\documentclass[]{aa} 

\usepackage{txfonts}
\usepackage{natbib}
\usepackage{graphics}
\usepackage{graphicx}
\usepackage{xcolor}
\usepackage{amssymb}
\usepackage{amsmath}
\usepackage{multicol}
\usepackage{subfig}

\bibpunct{(}{)}{;}{a}{}{,}
\begin{document}

   \authorrunning{Fangfang Song}
   \titlerunning{Binary open clusters in the Gaia data}
   
   \title{Binary open clusters in the Gaia data}

   \author{Fangfang Song\inst{1,2}, Ali Esamdin\inst{1,2},Qingshun Hu\inst{1,2}, Mengfan Zhang\inst{1,2}}
   
    \institute{Xinjiang Astronomical Observatory, Chinese Academy of Sciences, Urumqi, Xinjiang 830011, People's Republic of China;
    \email{aliyi@xao.ac.cn; songfangf@xao.ac.cn} \\
    University of Chinese Academy of Sciences, Beijing 100049, People's Republic of China;
    }

   \date{Received X XX, XXXX; accepted X XX, XXXX}

 
  \abstract
   {Observations indicate that the fraction of potential binary star clusters in the Milky Way is either the same or lower than that of the Magellanic Clouds. The unprecedented precision in the parallax measurements by Gaia has allowed for the discovery of a growing number of new binary open clusters (OCs).}
   {We aim to survey the candidates of truly binary open clusters that are  formed simultaneously, using information from the Gaia database.}
   {Based on the most recent catalog of open clusters, we investigated the interactions of adjacent binary open clusters in our Galaxy within separations of 50 pc. We compared their coordinates, proper motions, parallaxes, and color-magnitude diagrams (CMDs) via binary plots for all candidate pairs. The candidates of truly binary open clusters are selected on the basis of their common proper motions and consistent behaviors in the CMDs of different clusters that are limited to a separation of 50 pc.}
   {About ten pairs of the selected binary open clusters appear to be the same clusters, based on evidence that almost half of the cluster members are shared. Fourteen pairs are possibly true binaries, implying that they may come from the same clouds, among which five pairs are newly discovered. In addition, two clusters, UBC~46 and UBC~192, were found to be part of the stellar complex LISCA I. Our results confirm that OCs born in groups are usually composed of young open clusters.
}
   {}

   \keywords{open clusters and associations: general -- Astrometry}

\maketitle{}
%
\section{Introduction}\label{section:1}

Open clusters (OCs) are born in giant molecular clouds and can go on to form groups such as pairs, triplets, or higher multiplicity systems, according to observational evidence \citep{2016MNRAS.455.3126C}. Contrary to field stars, the distance, reddening, age, and metallicity for member stars of a given open cluster can be determined by the statistical cluster parameters \citep{2021A&A...649A..54P}. Among OCs, binary open clusters play a key role in understanding the formation, evolution, and properties of stars in the Galactic disk \citep{2021ARep...65..755C, 2021A&A...649A..54P}. These structures are also crucial in the study of the mechanisms that lead to cluster formation and evolution \citep{2018MNRAS.474.2277D}.

\citet{2009A&A...500L..13D} divided the formation of multiplicity star clusters into five different scenarios, including simultaneous formation, sequential formation, tidal capture, resonant trapping, and optical doubles. The clusters formed from the simultaneous formation are defined as genetic pairs or true binaries because they share common space velocities, ages, and chemical compositions. This means that either they come from the same molecular cloud or they are the final products of multiple mergers of smaller clusters\citep{2009A&A...500L..13D}. These bona fide pairs are ideal laboratories for promoting the development of formation and evolution theories of star clusters that are still poorly understood \citep{2022MNRAS.511L...1P}. 

The fraction of binary open clusters accounts for roughly 10\% of the known OCs in the Large Magellanic Cloud (LMC), according to the observations \citep{2009A&A...500L..13D}. The estimated fraction of binary clusters in our Galaxy is 12\% \citep{2009A&A...500L..13D}, a proportion that is similar to that of the LMC. Until recently, the binary open cluster of $h$ and $ \chi $ Persei has been the only confirmed genetic pair known in the Milky Way \citep{2010A&A...511A..38V}, which means that more binary open clusters have not been found in the disk.

\cite{1989SvA....33....6P} proposed the existence of five possible cluster groups. Later, \cite{1995A&A...302...86S} examined the spatial proximity of open clusters by the catalog of \cite{1995yCat.7092....0L} and suggested 18 probable binary open star clusters and found that about 8\% of open clusters may be genuine binaries. \cite{1997A&AT...14..181L} provided a catalog of 31 probable multiple systems by restricting the spatial vicinities and age coincidences. \citet{2009A&A...500L..13D} found 6 clear candidates of 34 OC pairs by constituted volume-limited samples from WEBDA \citep{2003A&A...410..511M} and NCOVOCC \citep{2002A&A...389..871D}. \cite{2017A&A...600A.106C} detected 19 groupings, including 14 pairs using an adapted version of the friends-of-friends algorithm on the 6D phase-space information, including radial velocities, for the first time, from the Catalogue of Open Cluster Data \citep{2004AN....325..740K, 2005A&A...440..403K} and the Radial Velocity Experiment \citep{2006AJ....132.1645S}. 

Thanks to the precisison of Gaia data, we have the opportunity to find more new open clusters and new binary open clusters. The second Gaia Data Release (Gaia DR2) provides full astrometric data (positions, parallaxes, and proper motions) and multi-band photometry for about 1.3 billion stars \citep{GaiaDR2}, leading to extensive studies of galactic OCs, namely, exclusions of false clusters, as well as accurate measurements of known star clusters, and searches for new clusters. \cite{Cantat2018} first determined a list of members and cluster parameters for 1229 clusters from the list of known clusters and candidates using Gaia DR2 data. These authors derived fully homogeneous parameters, including parallaxes, proper motions, and the most reasonable distances of the clusters, together with the membership probabilities of the individual stars (based on UPMASK). The number was later expanded to 1481 \citep{Cantat2020} and then updated to 2017 \citep[][CG20]{2020A&A...640A...1C} after a few months.

The steadily growing numbers of known open clusters and relatively accurate classification of cluster member stars based on Gaia DR2 has led to enthusiastic attempts to search for binary open clusters. \cite{Soubiran2019} provided 21 cluster pairs differing by less than 100 pc in distance and 5 km/s in velocity. \cite{2019ApJS..245...32L} found 56 candidates for star cluster groups among the Class 1 cluster candidates using the data of Gaia DR2 based on the star clusters’ 3D positions only. In the literature of \cite{2021RAA....21..117C}, a new binary system named Casado 9 and 10 was discovered in the reported 20 new OCs, which have been confirmed manually by Gaia DR2. There are 22 new binary or multiple OCs in a 30 deg sector of the Galactic disc found in \cite{2021ARep...65..755C}, based on the catalogues of OCs \citep{Kharchenko2013, Cantat2018, 2019AJ....157...12B, 2019ApJS..245...32L, 2019JKAS...52..145S, 2020A&A...635A..45C, 2020A&A...640A...1C}, identified by hand between the galactic longitudes of 240$^{\circ}$ and 270$^{\circ}$. Sixty aggregates of clusters are found in \cite{2021A&A...649A..54P} by searching for open clusters that\ share some member stars located at relatively low volumes of the phase space in the catalog of \cite{Cantat2020}. Nevertheless, the possibility for new discoveries of binary open clusters still exists. 

The aim of this work is to locate more genetic binary open clusters. The present study is organized as follows. In Section 2, we describe the methodology for the selection of candidate OCs belonging to groups and the analysis. Section 3 presents a discussion of the details of each OC  binary or multiple system and a comparison with the Two Micron All Sky Survey (2MASS) catalog \citep{2003yCat.2246....0C}. We summarize our results in Section 4.

\section{Pair selection and analysis}\label{section:2}

The data sets adopted in this work were derived by CG20, which is presented as two tables. The list of clusters (2017 entries) in their Table 1 includes mean positions, mean proper motions, and mean parallaxes, as well as other reliable parameters (ages, extinctions, distance modulus, distances converted from distance modulus, XYZ positions in Galactic cartesian coordinates, and distances from Galactic center) for 1867 clusters. A total of 234\,128 cluster member stars with probability over 0.7 are listed in their other table, which contains the astrometric measurements and photometric observational information for most of the stars by Gaia, the derived membership probabilities, and IDs of the host clusters. They employed uniform analysis methods to ensure unbiased comparisons among clusters.\citet{2019JPhCS1127a2053P} argued that systems located outside the dense region with small separations have a higher probability of being considered real pairs. The linking lengths used to identify pair open clusters are always limited to 100 pc \citep{2017A&A...600A.106C, 2019ApJS..245...32L}. \citet{2009A&A...500L..13D} argued that the pairs' physical separation must be less than 30 pc, while \citet{2022MNRAS.511L...1P} detected pairs with physical separations smaller than 40 pc. The basic criterion adopted in our preliminary selection is arbitrarily set to 50 pc (each pair's spatial separation) based on the 3D positions of the star clusters given in the datasets. The separation between two clusters is calculated using three-dimensional (3D) coordinates that include RA, DEC, and distance based on the Python Astropy package \citep{2013A&A...558A..33A, 2018AJ....156..123A}. The distance to each cluster is computed as $1/\omega$, with $\omega$ as the mean parallax of the cluster given by CG20.
   
We obtained a total of 125 preliminary pair candidates. Sixteen of them are found to contain some member stars, which have also been assigned to another cluster based on the same equatorial coordinates. This means that the host clusters may be physically close to each other -- or even overlapping \citep{2021A&A...649A..54P}. We assume that if the coincidence rate of the member stars in a cluster pair is greater than 50\%, they are considered to be one cluster; these results are listed in Table \ref{binaries1}. Table \ref{binaries1} provides the cluster names, the mean proper motions in RA and Dec, the mean parallaxes, the corresponding numbers of cluster members, the cluster ages, and the extinctions of the pair clusters in proper sequence (Columns 2-13), which are given by CG20. We calculate the cluster spatial separations and the star coincidence rates of cluster members listed in the last two columns of Table \ref{binaries1}. Most of the selected pairs are aggregates of clusters in Table A.1 of \citet{2021A&A...649A..54P}, except for numbers 8-10, which are not included in their paper since clusters UBC 167, UBC 392, and UBC 323 are newly found clusters in the work of \citet{Cantat2020}. Among the ten pairs, the star coincidence rates of four pairs are larger than 0.8, which means they have a greater probability of being the same clusters. The star coincidence rates of the other four pairs are approximately 0.7 and the values of the other two pairs are closer to 0.5. These pairs with star coincidence rates lower than 0.8 require more evidence to verify their characterization. 

After eliminating those pairs (as shown in Table \ref{binaries1}), details such as coordinates, proper motions, parallaxes, and color-magnitude diagrams (CMDs) can be checked by visual inspection, as shown in Figure \ref{fig:fig1}. If the proper motions of two clusters are similar in panel a and consistent within 3 $\sigma$ error range (as shown in Table \ref{binaries2}), then we considered them as having common proper motions. According to \citet{2020A&A...642L...4K}, if the ages and reddenings of the two clusters cannot be distinguished in the CMD, then the cluster ages and reddenings seem to be equal and the same values can be used for both clusters. Furthermore, the indistinguishable CMDs (panel b) indicates similar ages. The differences in coordinates or parallaxes (panels c, d) indicate that the clusters can be distinguished in the 3D space. The pairs exhibiting similar proper motions and ages can be considered as candidates of truly binary open clusters. 

In total, we obtained 14 candidate truly pairs from the limited 115 candidate pairs, which are given in Table \ref{binaries2}. The listed parameters are the same as those in Table \ref{binaries1}. Most of the proper motions are consistent in the error range for each pair of the 14 candidates, except pair 5, whose proper motions in the declination direction are consistent within the 2 $\sigma$ error range. The distributions of proper motions, CMDs, parallaxes, and positions of members for the selected 14 open cluster pairs are shown in Figure \ref{fig:fig1}. All the adopted parameters of Figure \ref{fig:fig1} were taken from Gaia DR3 \citep{2022yCat.1355....0G} by cross-matching the equatorial coordinates, because the photometric and astrometric uncertainties are greatly improved compared with those in Gaia DR2. The dark dots and gray histograms represent the former clusters, while the red dots and red histograms are related to the latter clusters. The black and red lines in CMDs are the fitted PARSEC stellar isochrones \citep{2012MNRAS.427..127B}, with the cluster metallicities assumed to be solar. Most isochrones are aptly fit with the parameters of CG20 of these candidates. In contrast, cluster parameters need to change to fit the member distributions, which are the clusters marked with asterisks in Table \ref{binaries2}. The cluster ages and extinctions of CG20 are approximated by artificial neural networks (ANN) with Gaia photometry and parallaxes, and the precision of ANN is affected by the numbers of cluster member stars and the densities of clusters (CG20). Therefore, some cluster isochrone fittings are reconfirmed in this work. Because they are assumed as part of binary open clusters, we can use the same ischrones of their partners. Pairs 2, 3, and 10 are aptly fit with the parameters of latter clusters, whereas Pair 4 and 12 are aptly fit with the parameters of former clusters. The revised parameters can be seen in Table \ref{binaries3}. The clusters marked with an asterisk have some common characteristics, namely, they contain fewer than 100 members and most of them have sparse distributions for their member stars in the CMDs, with their spacial positions shown in Figure \ref{fig:fig1}. In addition, two clusters, UBC~46 and UBC~192, were found to be part of the $h$ and $ \chi $ Persei double clusters in our study (see details in Section 3.1.1).

\begin{table*}
\centering
\fontsize{8} {10pt}\selectfont
\tabcolsep 0.10truecm
\caption{List of possible one clusters}
\begin{tabular}{rccrrrrrrrrrrrrrr}
\hline\hline
\multicolumn{1}{c}{No. \#}         &
\multicolumn{1}{c}{Cluster}       &
\multicolumn{1}{c}{$\mu_{\alpha}\cos\delta$}       &
\multicolumn{1}{c}{$\mu_{\delta}$}       &
\multicolumn{1}{c}{$\omega$}       &
\multicolumn{1}{c}{members}      &
\multicolumn{1}{c}{$\tau$}       &
\multicolumn{1}{c}{Av}       &
\multicolumn{1}{c}{Sep}      &
\multicolumn{1}{c}{ratio}      &  \cr
& &(mas yr$^{-1}$) & (mas yr$^{-1}$) &(mas) &   &(Myr) & (mag) &(pc) &  \cr 

\hline
1 & BH 121       & -6.0(1) & 0.6(1)   & 0.39(4)  & 174 & 2.63    & 1.11 & 7.78  & 0.92 \\
  & IC 2948      & -6.0(1) & 0.65(5)  & 0.39(3)  & 51  & 6.46    & 1.16 &       &     \\
2 & FSR 0686     & -1.1(1) & -2.56(7) & 1.09(4)  & 12  &         &      & 6.83  & 0.91 \\
  & UBC 55       & -1.1(1) & -2.6(1)  & 1.08(5)  & 56  & 33.88   & 1.38 &       &     \\
3 & RSG 7        & 4.9(4)  & -1.9(7)  & 2.3(1)   & 43  & 38.90   & 0.57 & 24.78 & 0.72\\
  & RSG 8        & 5.3(5)  & -1.7(5)  & 2.2(2)   & 39  & 26.92   & 0.53 &       &     \\
4 & Gulliver 56  & 0.53(8) & -3.24(7) & 0.46(4)  & 18  & 407.38  & 0.44 & 48.74 & 0.83 \\
  & UBC 73       & 0.5(1)  & -3.21(9) & 0.45(3)  & 50  & 389.05  & 0.56 &       &     \\
5 & Gulliver 6   & -0.0(4) & -0.2(4)  & 2.4(1)   & 318 & 16.60   & 0.25 & 1.78  & 0.72 \\
  & UBC 17b      & 0.1(1)  & -0.2(3)  & 2.38(5)  & 103 & 11.48   & 0.05 &       &     \\
6 & Hogg 10      & -6.20(7)& 1.75(5)  & 0.38(2)  & 11  &         &      & 41.99 & 0.72 \\
  & NGC 3572     & -6.3(1) & 1.9(2)   & 0.38(5)  & 75  & 4.79    & 1.53 &       &     \\
7 & Kronberger 1 & -0.1(1) & -2.2(2)  & 0.4(1)   & 32  & 6.03    & 2.05 & 20.29 & 0.53\\
  & Stock 8      & 0.1(1)  & -2.2(2)  & 0.45(5)  & 275 & 14.45   & 1.17 &       &     \\
8 & UBC 10a      & -2.1(1) & -3.0(1)  & 1.08(2)  & 33  & 14.13   & 0.89 & 9.03  & 0.56 \\
  & UBC 167      & -2.08(9)& -3.0(1)  & 1.07(3)  & 115 & 21.88   & 1.08 &       &     \\
9 & UBC 392      & -0.6(1) & -3.3(1)  & 1.06(3)  & 55  & 70.79   & 1.47 & 8.55  & 0.96\\
  & UPK 194      & -0.6(2) & -3.1(3)  & 1.05(5)  & 267 & 151.36  & 1.67 &       &     \\
10& BH 205       & -0.2(1)  & -1.1(2) & 0.57(7)  & 124 & 6.17    & 1.23 & 18.74 & 0.69\\
  & UBC 323      & -0.2(2) & -1.3(2)  & 0.58(5)  & 170 & 8.91    & 0.72 &       &     \\
\hline
\end{tabular}
\begin{list}{}{}
{\footnotesize
  \item[] \textbf{Notes} $\mu_{\alpha}\cos\delta$,$\mu_{\delta}$: average proper motions of cluster members in RA and Dec mas yr$^{-1}$; $\omega$:Mean parallax of cluster members in mas;$members$: Numbers of cluster member stars; $\tau$: cluster age in Myr ; $Av$: Extinction of the cluster in mag; All the above parameters were taken from CG20. Sep: cluster pairs' spatial separation in pc; ratio: the star coincidence rate of cluster members. This Sep and ratio values were calculated in this work.

}
\end{list}
\label{binaries1}
\end{table*}

\begin{table*}
\centering
\fontsize{8} {10pt}\selectfont
\tabcolsep 0.10truecm
\caption{List of candidates of truly binary open clusters}
\begin{tabular}{rccrrrrrrrrrrrrrr}
\hline\hline
\multicolumn{1}{c}{Pairs \#}         &
\multicolumn{1}{c}{Cluster}       &
\multicolumn{1}{c}{$\mu_{\alpha}\cos\delta$}       &
\multicolumn{1}{c}{$\mu_{\delta}$}       &
\multicolumn{1}{c}{$\omega$}       &
\multicolumn{1}{c}{members}      &
\multicolumn{1}{c}{$\tau$}       &
\multicolumn{1}{c}{Av}       &
\multicolumn{1}{c}{Sep}      &   \cr
& &(mas yr$^{-1}$) & (mas yr$^{-1}$) &(mas) &   &(Myr) & (mag) &(pc)   \cr 

\hline
1 & NGC 869             & -0.7(1) & -1.1(1) & 0.40(4) & 503 & 15.14 & 1.69 & 21.01\\
  & NGC 884             & -0.6(1) & -1.1(1) & 0.40(4) & 322 & 17.78 & 1.72 &      \\
2 &$UBC~46^{\star}$     & -0.8(2) & -1.2(2) & 0.40(3) & 65  &112.20 & 1.35 & 29.30\\
  & NGC 869             & -0.7(1) & -1.1(1) & 0.40(4) & 503 & 15.14 & 1.69 &      \\
3 &$UBC 192^{\star}$    & -0.4(2) & -1.2(1) & 0.40(3) & 78  & 48.98 & 1.28 & 38.59\\
  & NGC 884             & -0.6(1) & -1.1(1) & 0.40(4) & 322 & 17.78 & 1.72 &      \\
4 &UBC 461              & -2.7(1) & 3.11(8) & 0.29(2) & 72  & 47.86 & 0.45 & 27.12 \\
  & $UBC~462^{\star}$   & -2.67(8)& 3.04(4) & 0.29(2) & 8   & 39.81 & 0    &      \\
5 &Collinder 135        & -10.0(4)& 6.2(4)  & 3.3(1)  & 324 & 26.30 & 0.01 & 26.42\\
  &UBC 7                & -9.7(2) & 7.0(2)  & 3.56(6) & 77  & 31.62 & 0.1  &      \\
6 &NGC 2659             & -5.3(1) & 5.0(1)  & 0.45(7) & 82  & 43.65 & 1.21 & 49.45\\
  & UBC 482             & -5.2(1) & 5.0(2)  & 0.45(3) & 97  & 26.92 & 0.88 &      \\
7 &FSR 0198             & -3.6(2) & -6.6(2) & 0.49(5) & 82  & 4.68  & 2.49 & 28.38\\
  & Teutsch 8           & -3.5(1) & -6.7(3) & 0.49(7) & 28  & 3.98  & 2.2  &      \\
8 &Collinder 394        & -1.5(2) & -5.9(2) & 1.39(6) & 156 & 91.20 & 0.54 & 11.49\\
  & NGC 6716            & -1.5(2) & -6.0(2) & 1.41(8) & 76  & 97.72 & 0.43 &      \\
9 &Alessi 43            & -5.5(3) & 3.9(3)  & 1.03(8) & 284 & 11.48 & 0.99 & 19.76\\
  & Collinder 197       & -5.8(3) & 3.9(4)  & 1.03(9) & 229 & 14.13 & 1.42 &      \\
10&$FSR 1297^{\star}$   & -2.8(2) & 3.3(2)  & 0.75(5) & 25  &       &      & 41.13\\
  & NGC 2362            & -2.8(2) & 3.0(2)  & 0.74(7) & 144 & 5.75  & 0.37 &      \\
11&Antalova 2           & 2.0(3)  & -1.7(3) & 0.84(7) & 27  & 12.02 & 1.15 & 47.02\\
  & NGC 6383            & 2.6(3)  & -1.7(2) & 0.9(1)  & 245 & 3.98  & 1.3  &      \\
12&Negueruela 1         & -3.0(2) & -1.3(1) & 0.29(6) & 52  & 16.98 & 3.39 & 18.31\\
  & $Teutsch~23^{\star}$& -3.0(1) & -1.39(7)& 0.29(4) & 28  & 5.62  & 2.69 &      \\
13& UBC 547             & -0.7(1) & -2.0(1) & 0.31(3) & 175 & 15.14 & 1.63 & 23.54\\
  & UBC 549             & -0.81(9)& -1.93(8)& 0.31(2) & 20  & 85.11 & 1.92 &      \\
14& Stock 20            & -3.24(6)& -1.12(6)& 0.34(3) & 38  & 19.05 & 1.09 & 23.74\\
  & UBC 410             & -3.1(1) & -1.0(1) & 0.34(3) & 36  & 38.02 & 0.97 &      \\
\hline
\end{tabular}
\begin{list}{}{}
{\footnotesize
  \item[] \textbf{Notes} 
  The given parameters are the same as those in Table \ref{binaries1}. The fitting isochrones of the clusters marked with asterisks are not suitable for the cluster members, which require an adjustment of the parameters to find adequate fittings. 
}
\end{list}
\label{binaries2}
\end{table*}

\begin{figure*}[!]
   \centering
   \begin{tabular}{ll}
   \subfloat{\includegraphics[scale=0.5]{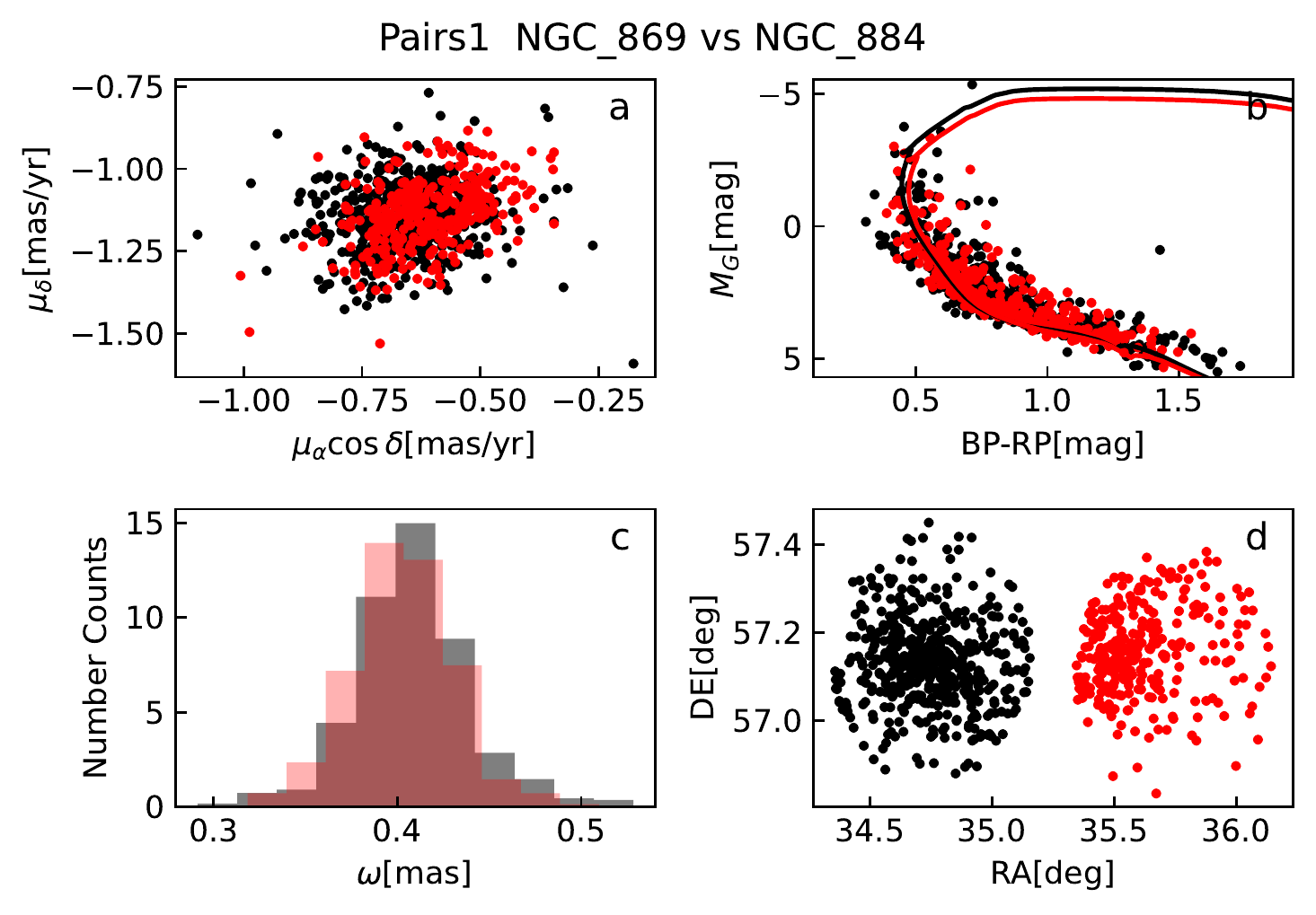}} \quad & 
   \subfloat{\includegraphics[scale=0.5]{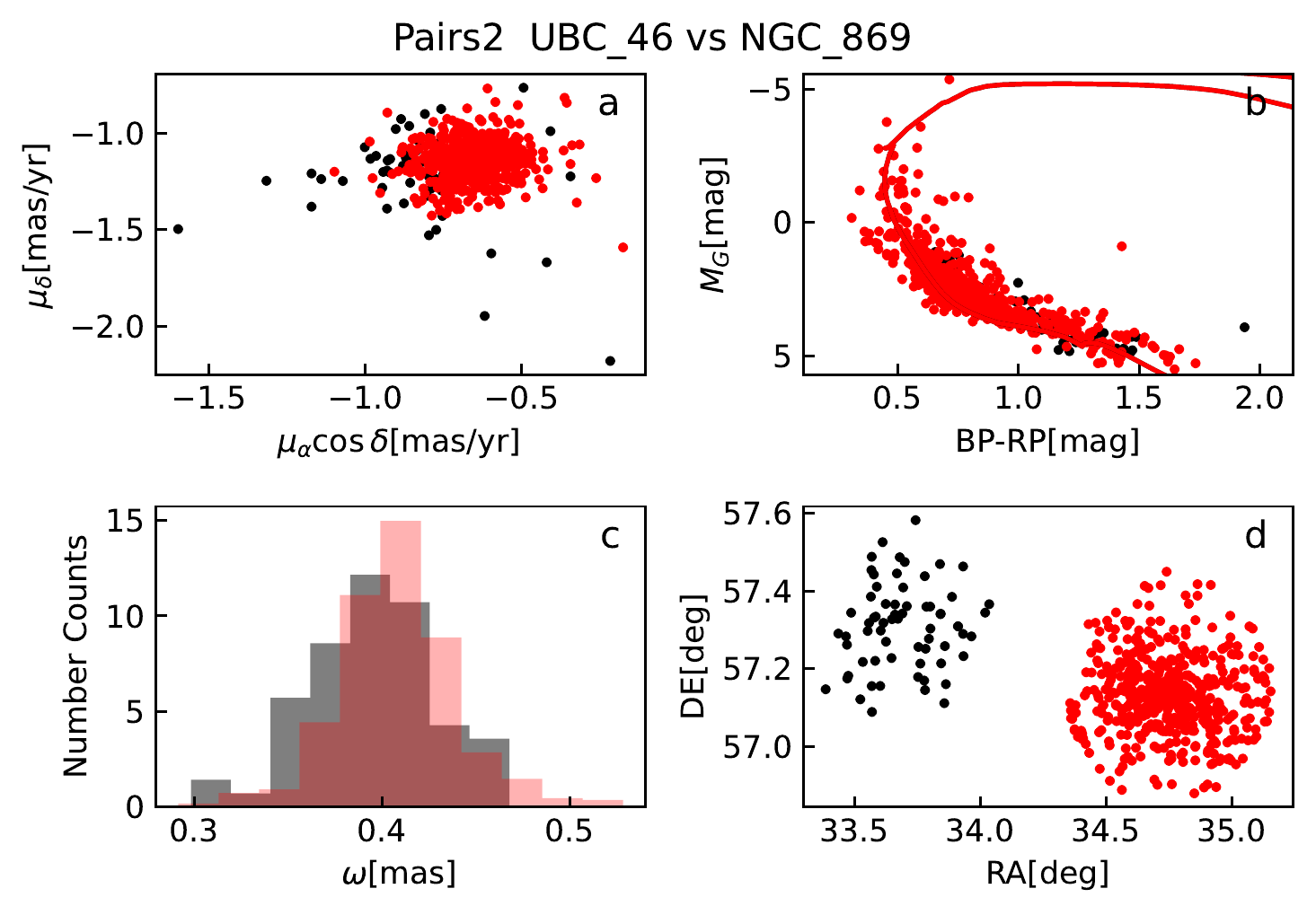}} \\
   \subfloat{\includegraphics[scale=0.5]{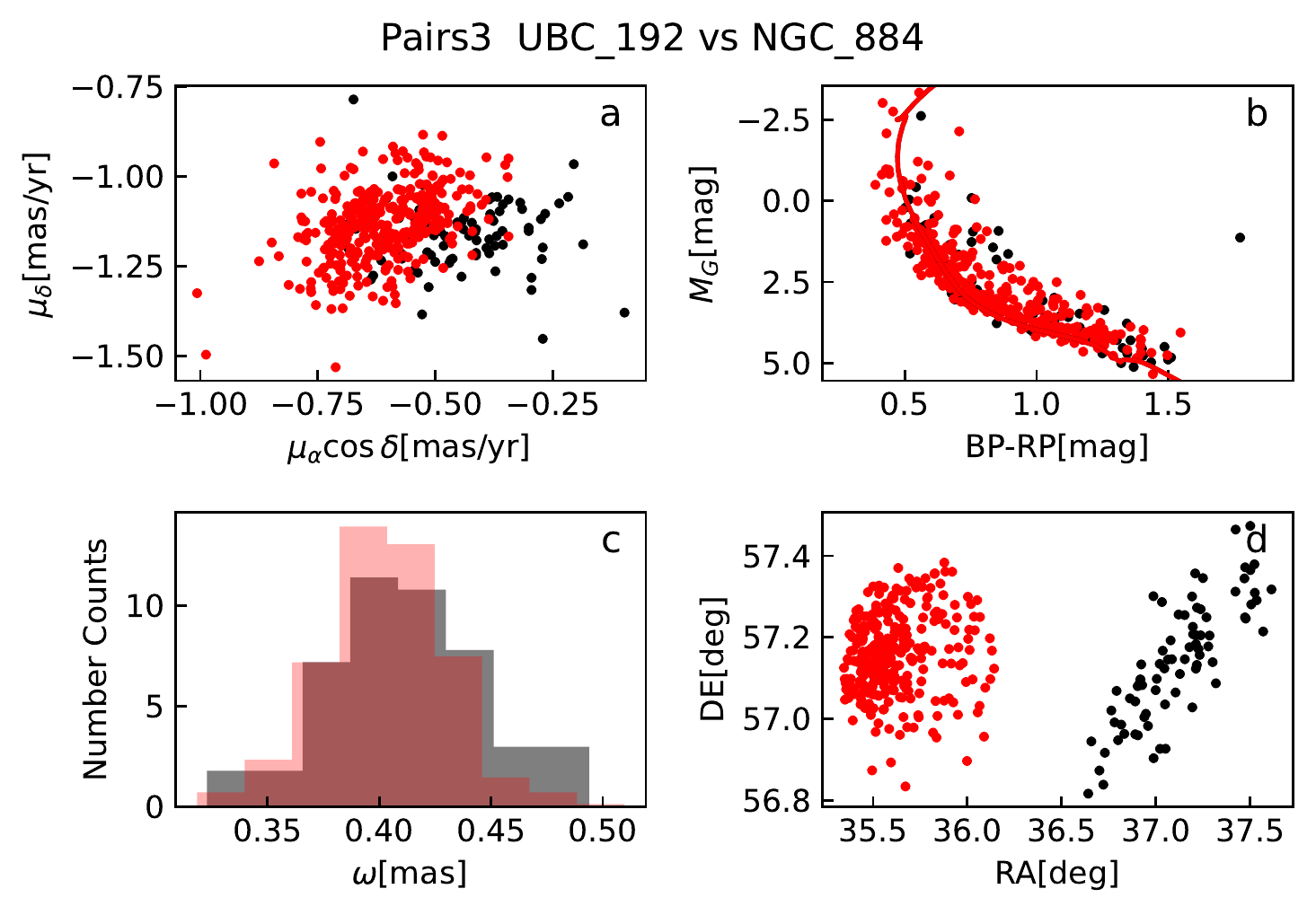}} \quad &
   \subfloat{\includegraphics[scale=0.5]{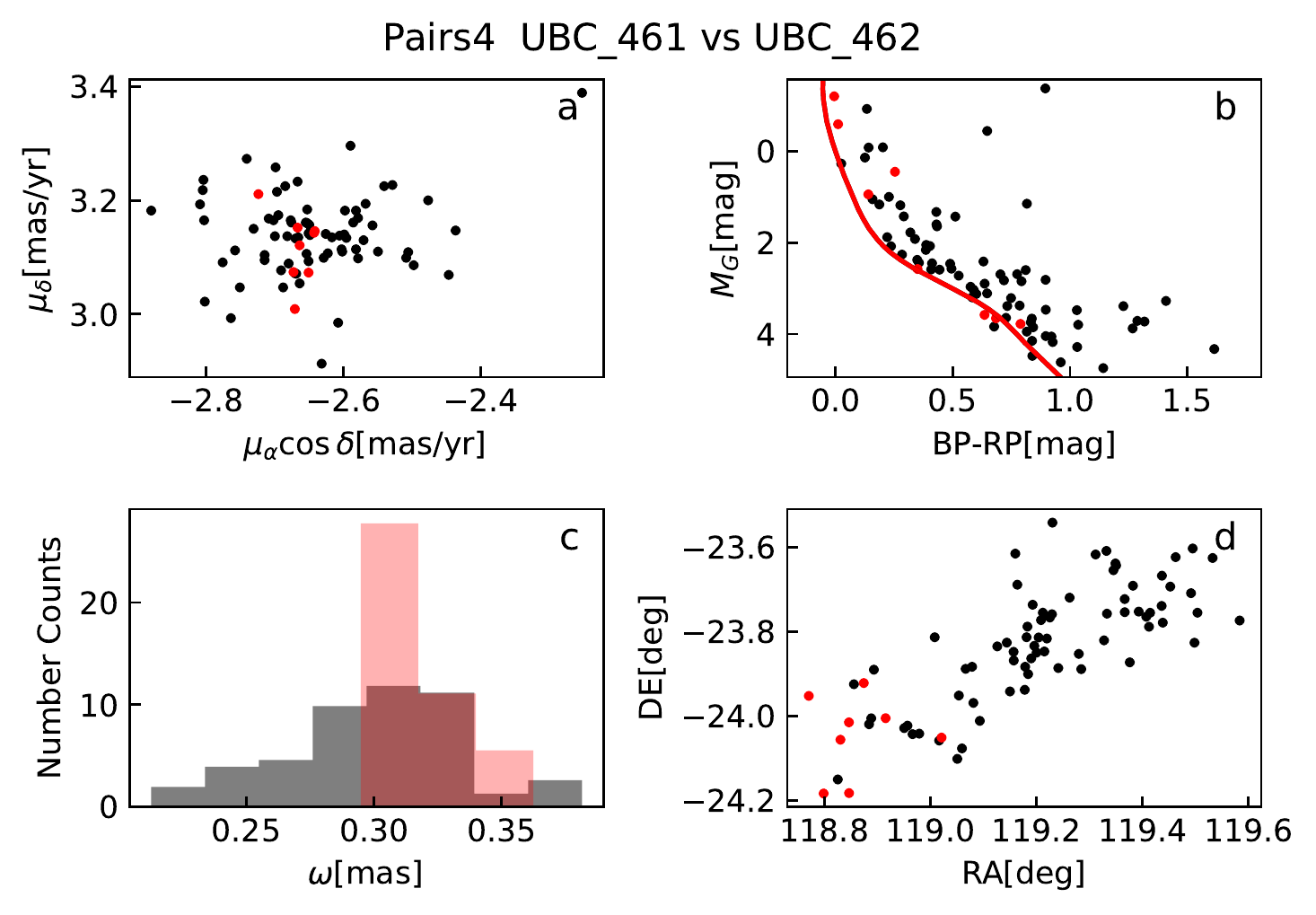}} \\
   \subfloat{\includegraphics[scale=0.5]{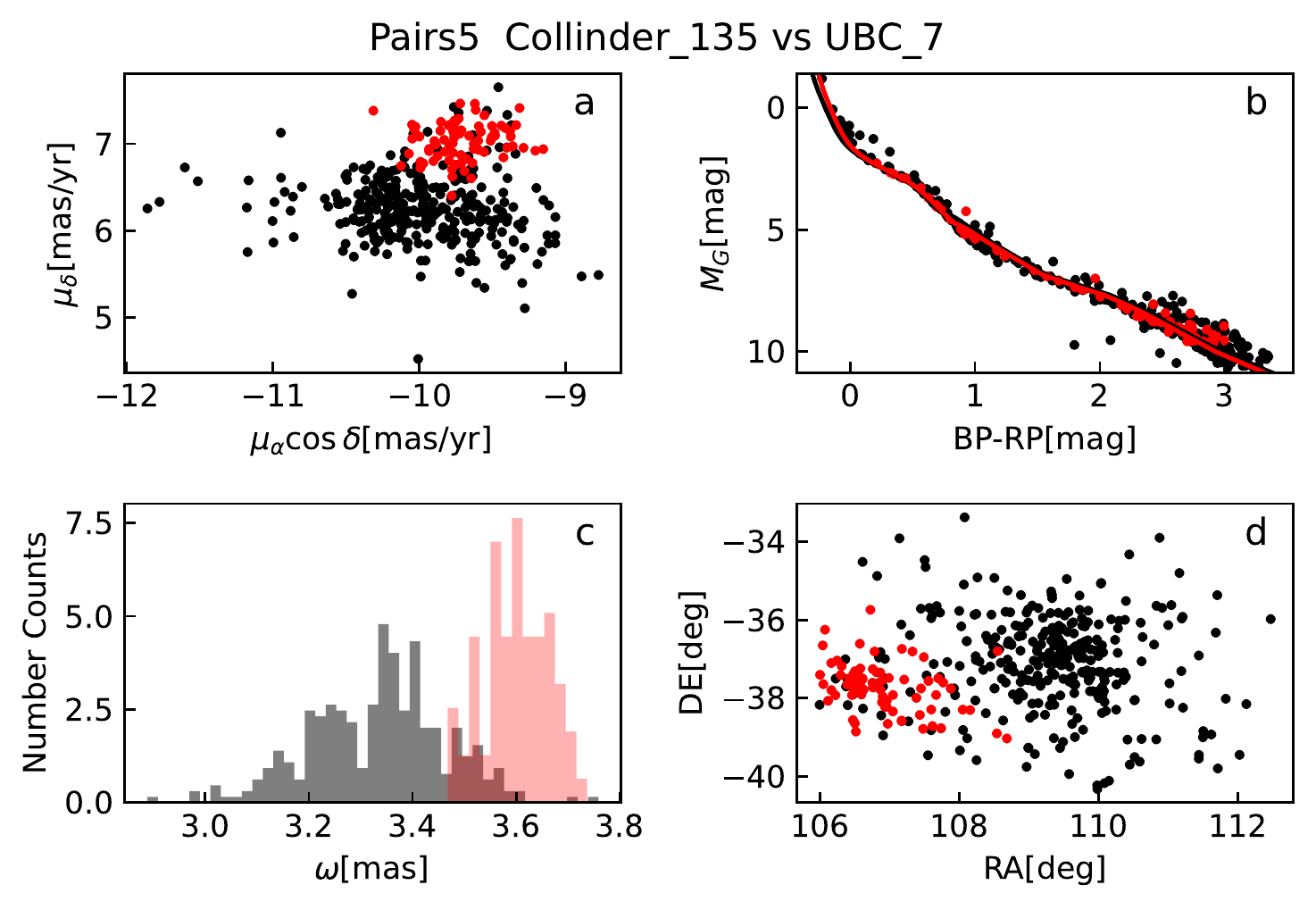}} \quad & 
   \subfloat{\includegraphics[scale=0.5]{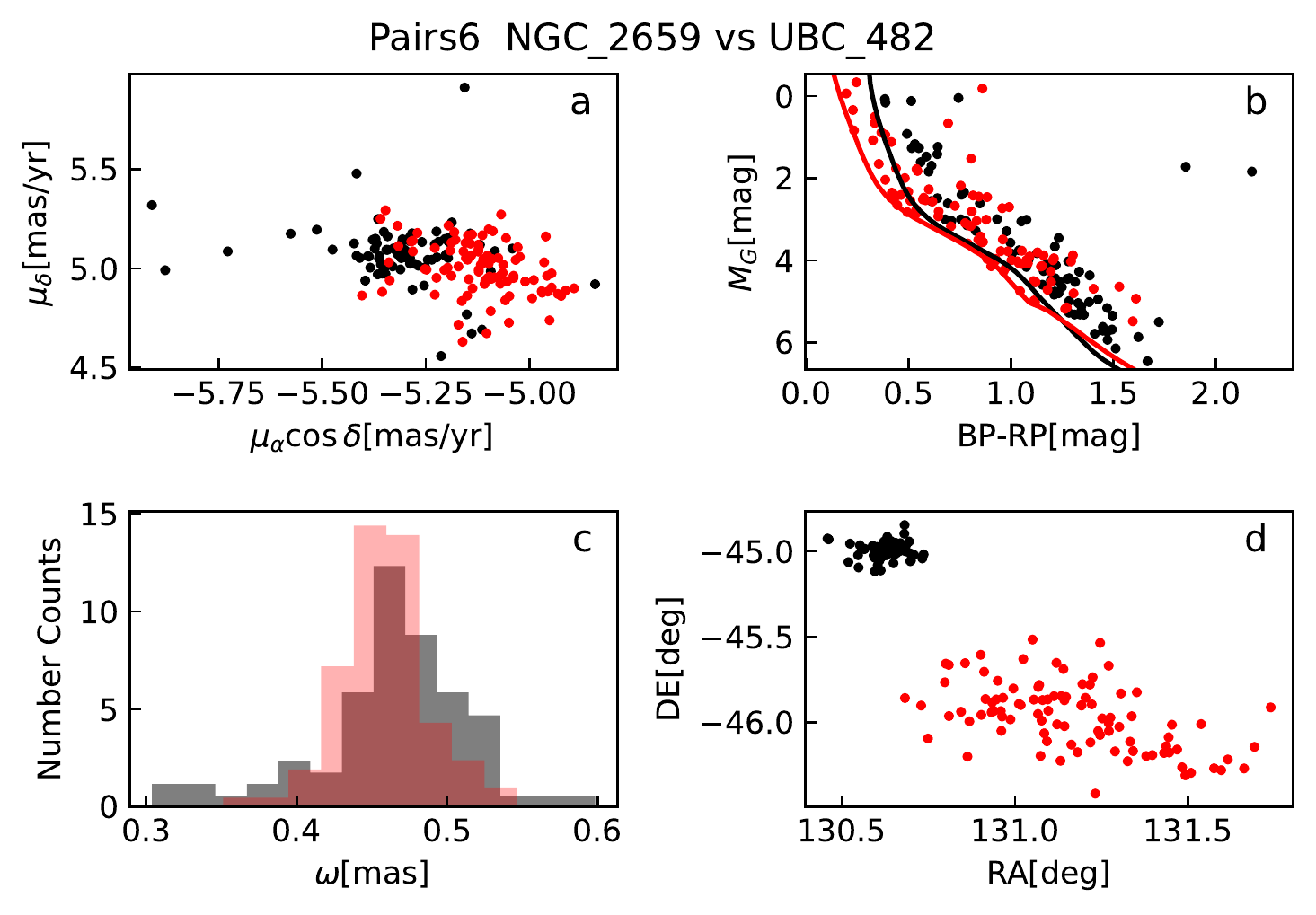}} \\
   \subfloat{\includegraphics[scale=0.5]{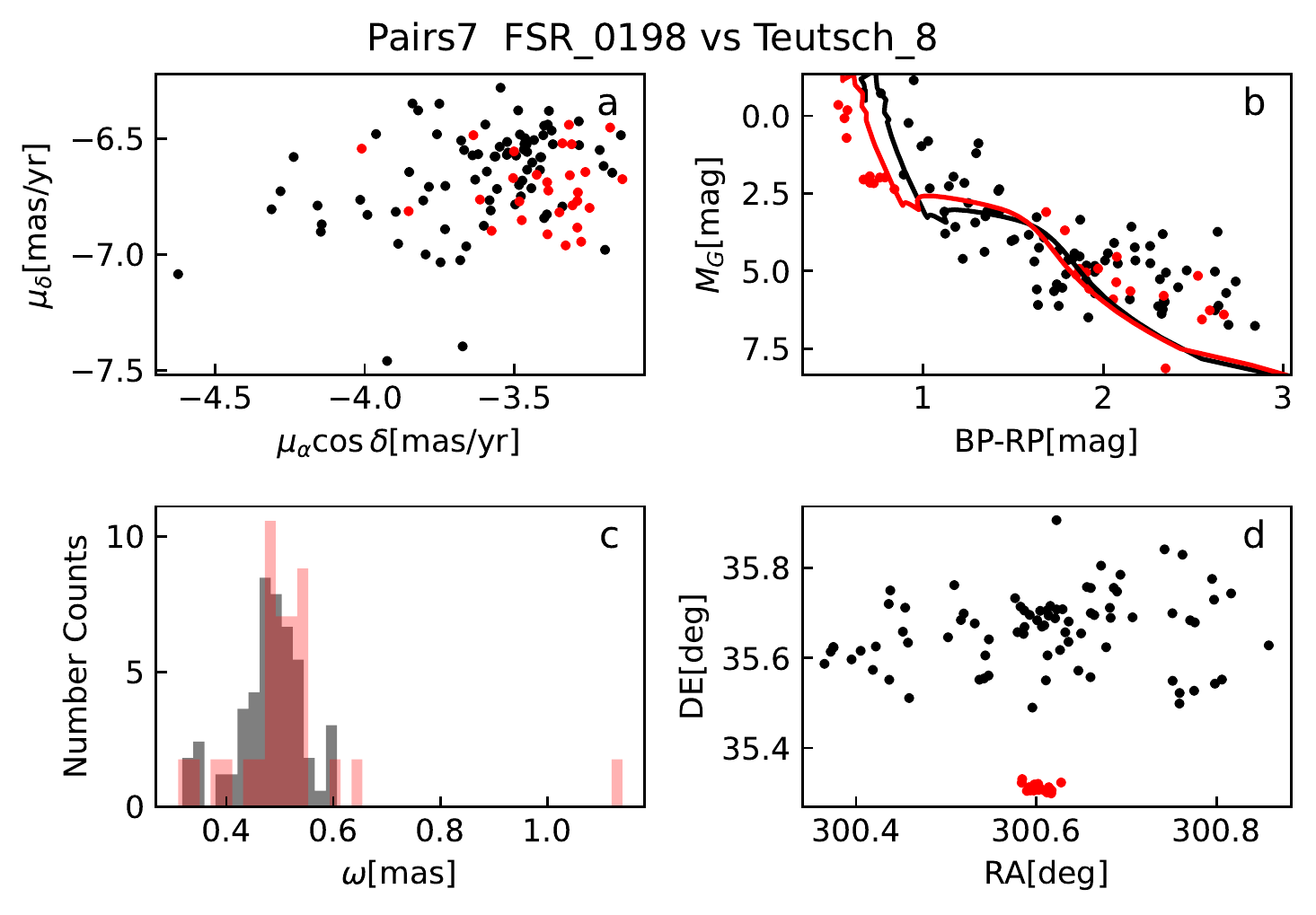}} \quad & 
   \subfloat{\includegraphics[scale=0.5]{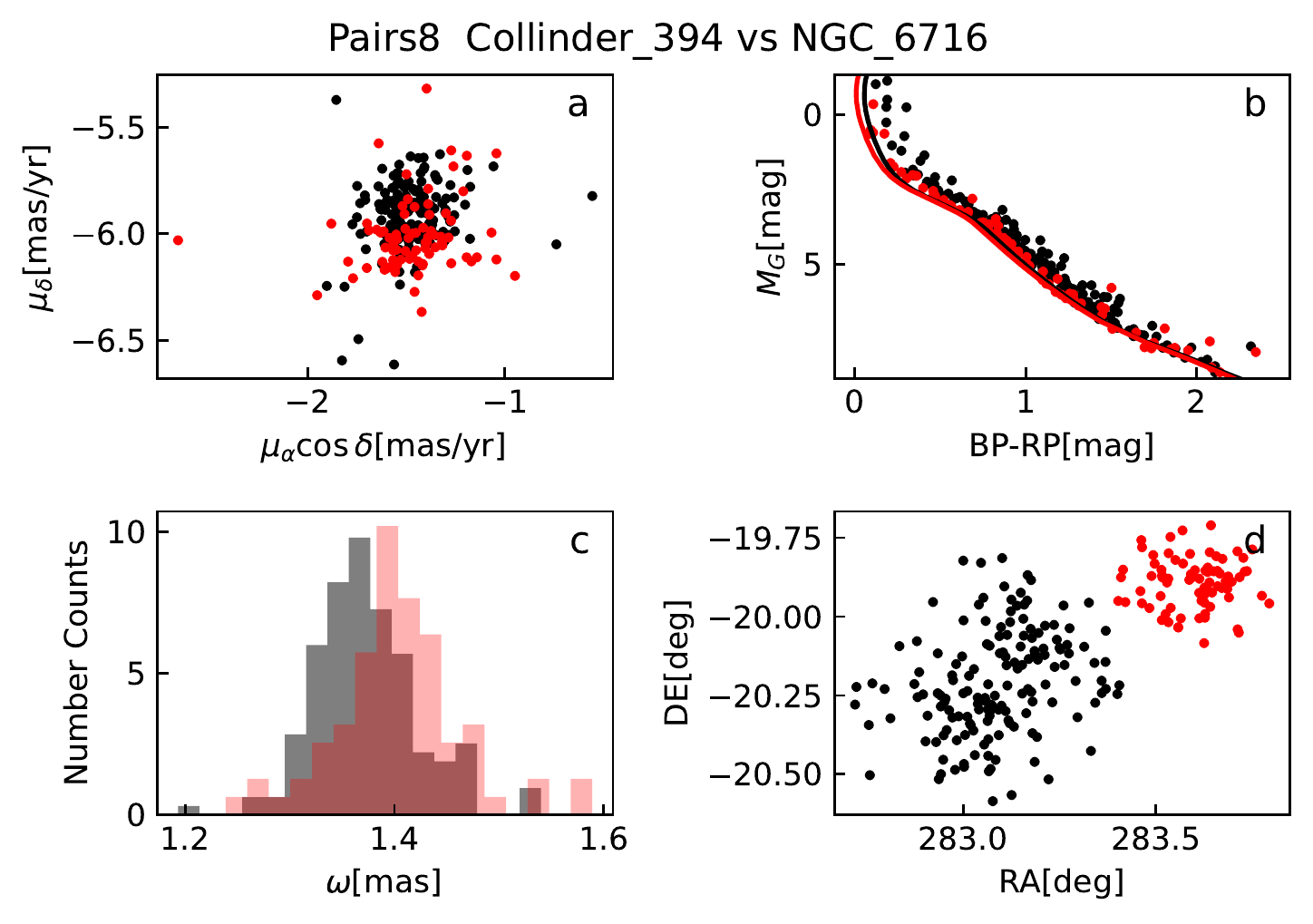}} \\     
   \end{tabular}
   \caption{Distributions of (a) proper motions, (b) CMDs, (c) parallaxes, and (d) celestial positions of member stars of selected open cluster pairs. The dark dots and gray histograms represent the former clusters, while the red dots and red histograms signify the latter clusters. The black and red lines in CMDs are the fitted PARSEC stellar isochrones.}
   \label{fig:fig1}
\end{figure*}
\begin{figure*}[!]
   \addtocounter{figure}{-1}
   \centering
    \begin{tabular}{ll}   
   \subfloat{\includegraphics[scale=0.5]{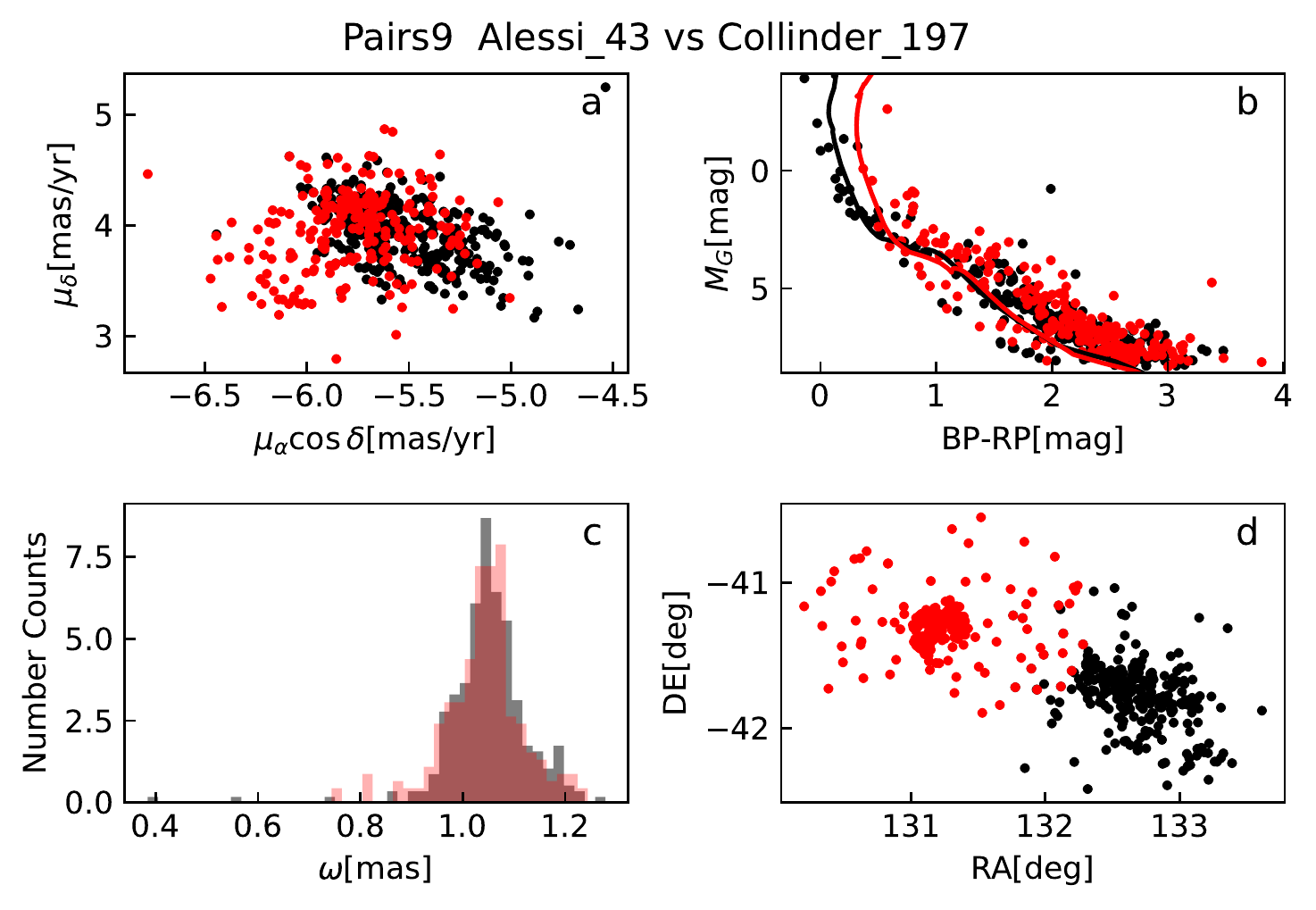}} \quad &
   \subfloat{\includegraphics[scale=0.5]{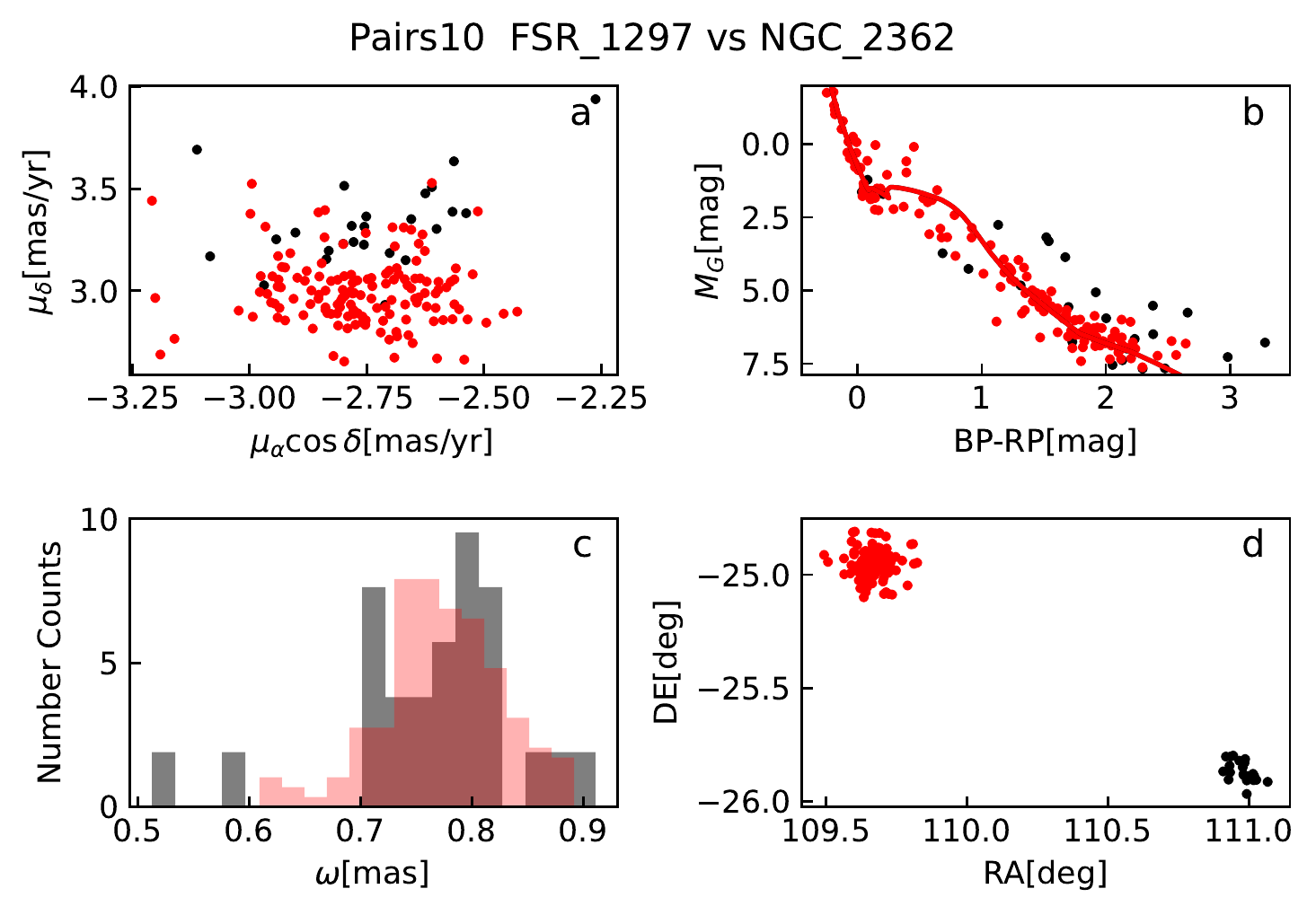}} \\ 
   \subfloat{\includegraphics[scale=0.5]{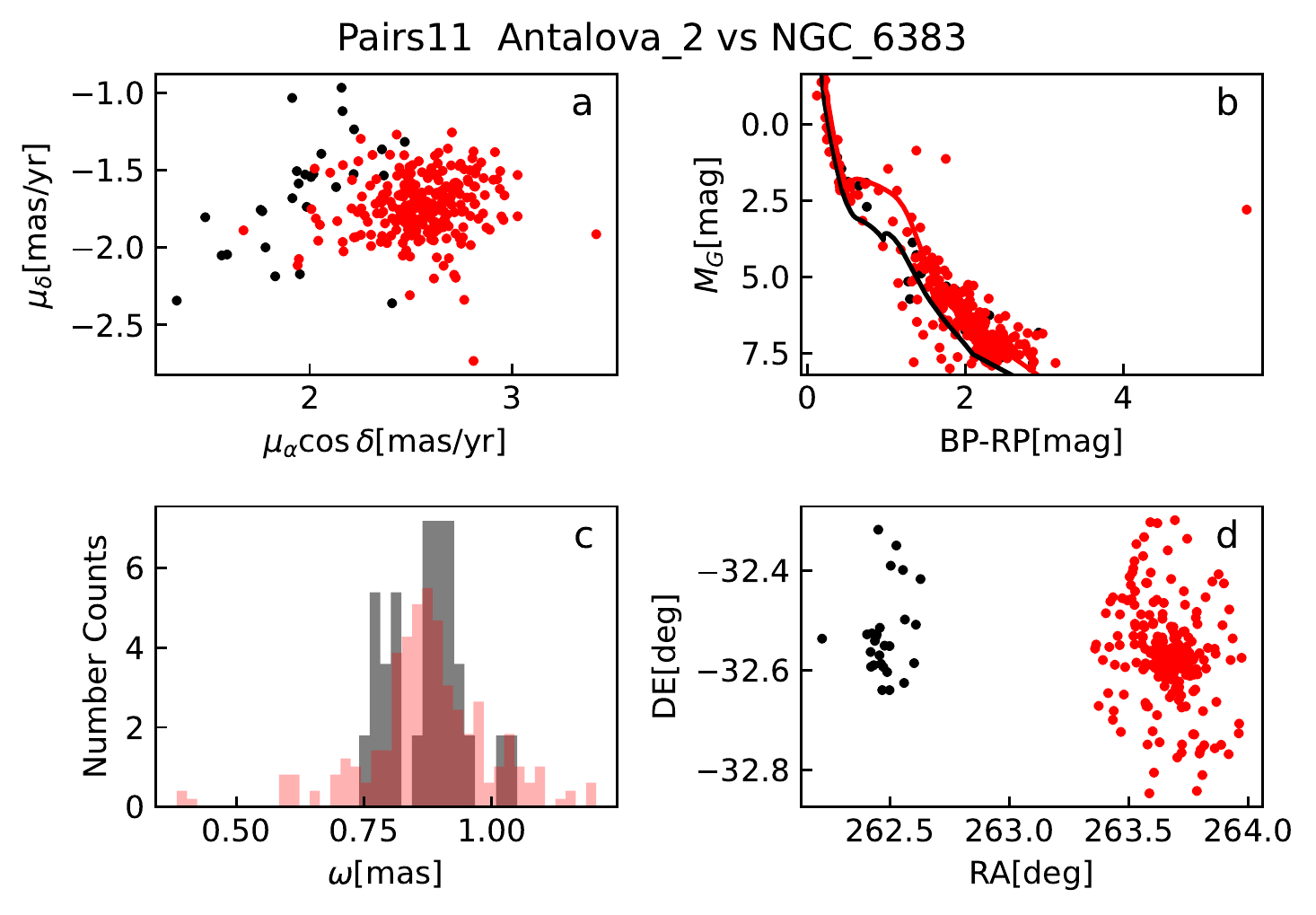}} \quad & 
   \subfloat{\includegraphics[scale=0.5]{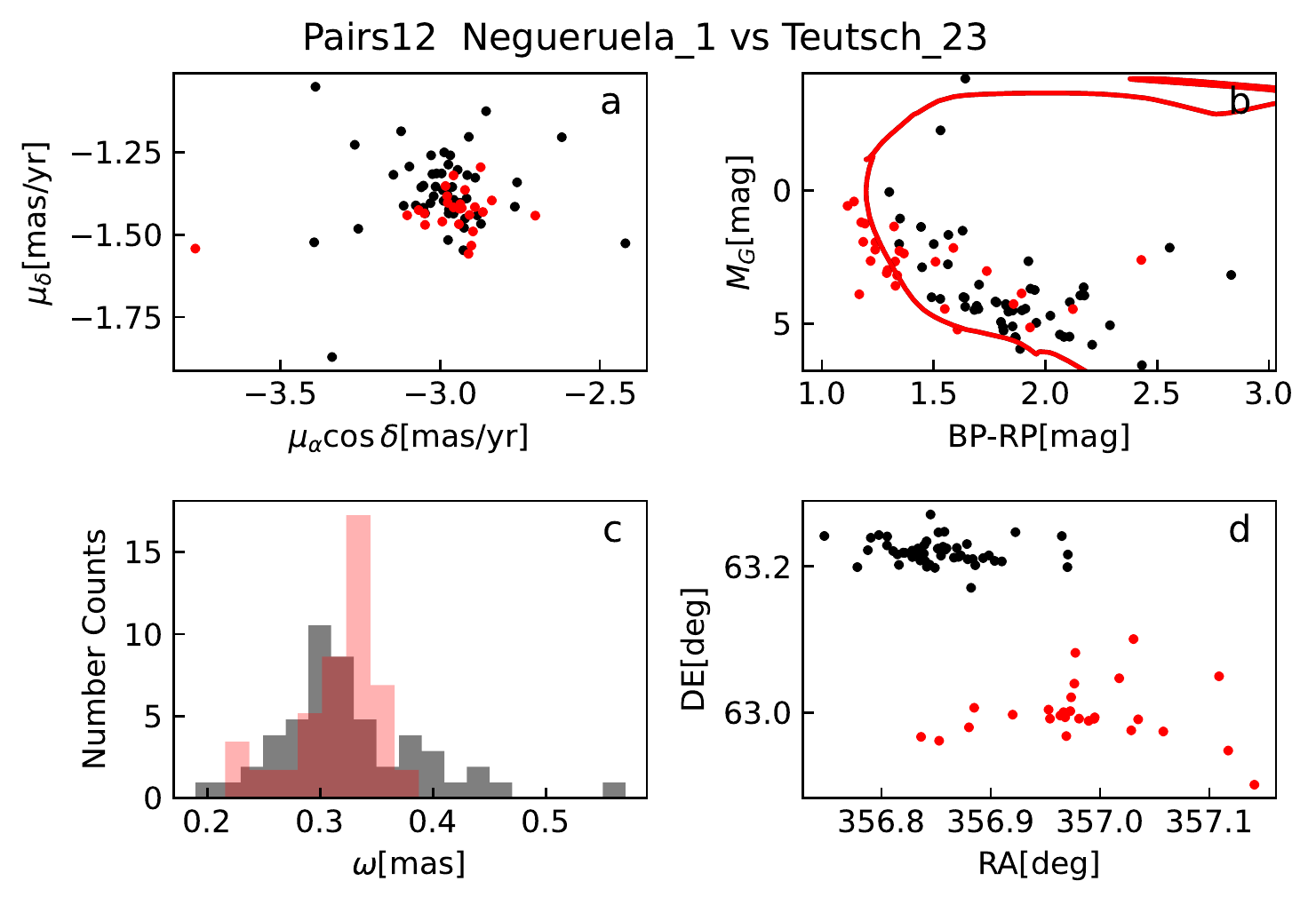}} \\
   \subfloat{\includegraphics[scale=0.5]{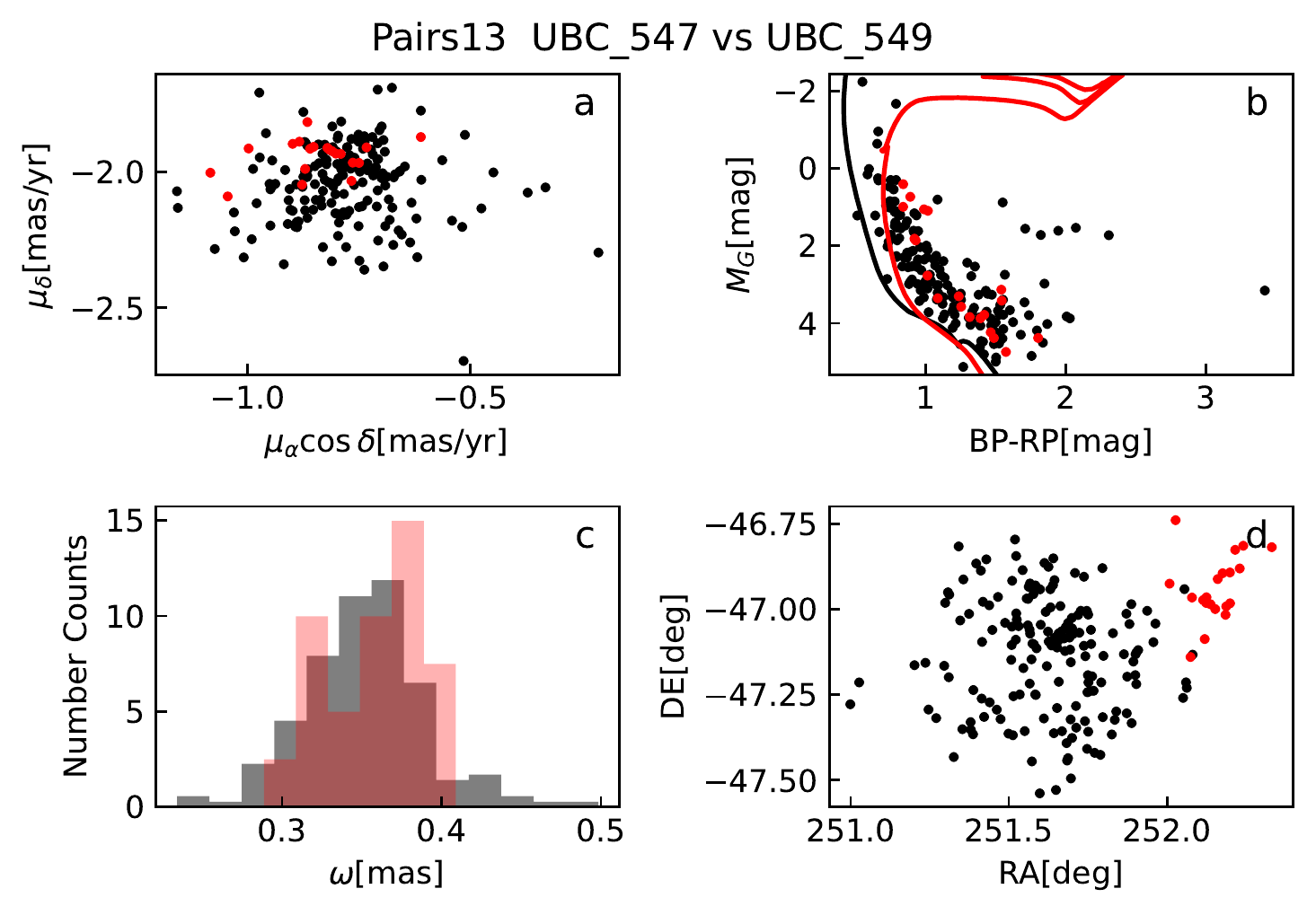}} \quad & 
   \subfloat{\includegraphics[scale=0.5]{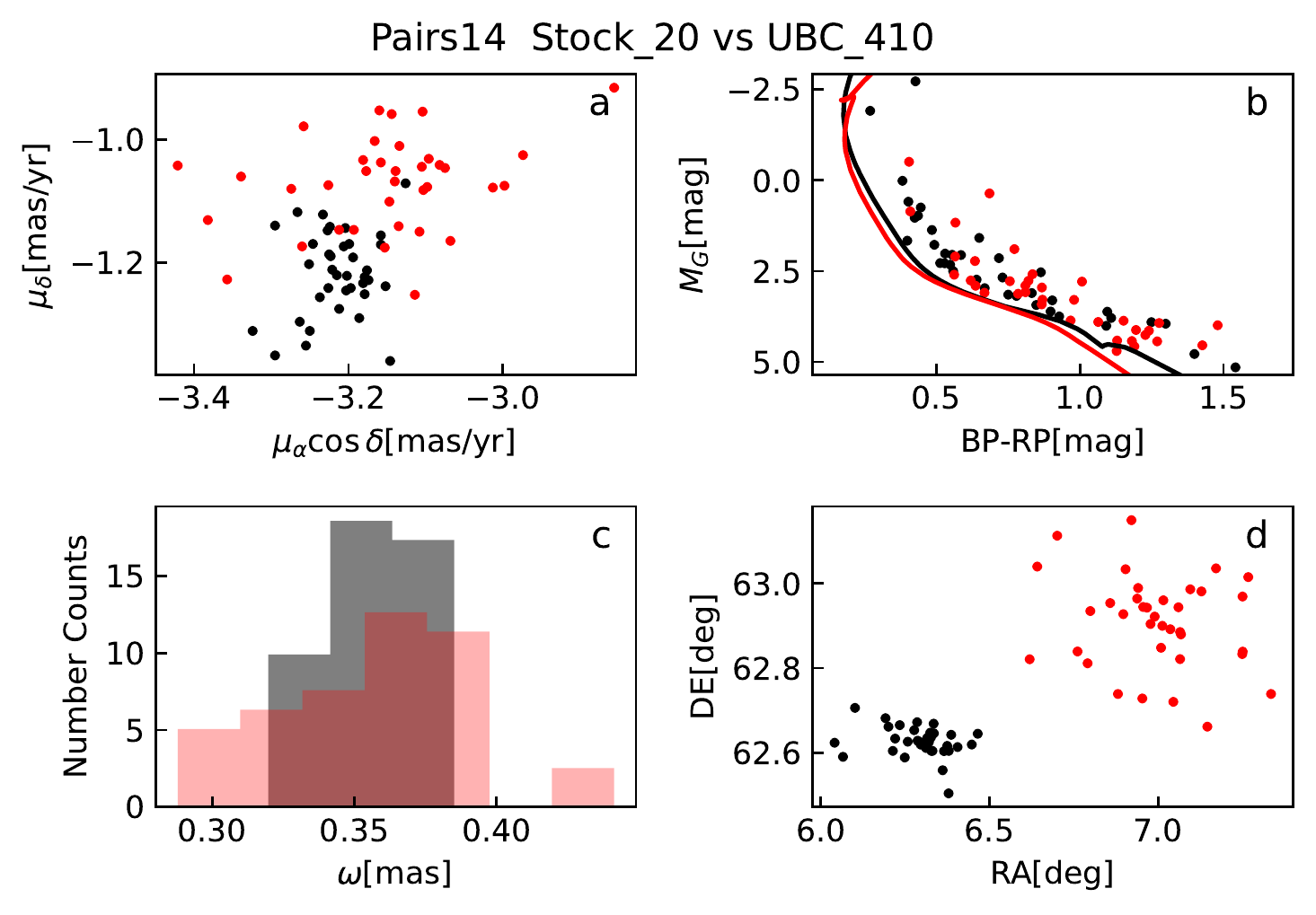}} \\
      
         \end{tabular}
         \caption{Cont.}
\end{figure*}

\begin{table*}
\centering
\fontsize{8} {10pt}\selectfont
\tabcolsep 0.10truecm
\caption{Comparation of the fitting parameters: ages and extinctions  of candidates for truly binary open clusters}
\begin{tabular}{rccrrrrrrrrrrrrrr}
\hline\hline
\multicolumn{1}{c}{pairs \#}         &
\multicolumn{1}{c}{Cluster}       &
\multicolumn{2}{c}{CG20}       &
\multicolumn{2}{c}{DAML02}       &
\multicolumn{2}{c}{MWSC}       &
\multicolumn{2}{c}{This work}       \cr\cline{3-10}
& &log(t) (yr) & $Av$ (mag) &log(t) (yr) & $Av$ (mag) &log(t) (yr) & $Av$ (mag) &log(t) (yr) & $Av$ (mag) \cr 

\hline
1-3 &NGC 869          & 7.18 & 1.69 & 7.07 & 1.78 & 7.28 & 1.62 & 7.18 & 1.69  \\
    &NGC 884          & 7.25 & 1.72 & 7.10 & 1.74 & 7.20 & 1.74 & 7.25 & 1.72 \\
    &$UBC~46^{\star}$ & 8.05 & 1.35 &      &      &      &      & 7.18 & 1.69\\
    &$UBC 192^{\star}$& 7.69 & 1.28 &      &      &      &      & 7.25 & 1.72\\
4   &UBC 461          & 7.68 & 0.45 &      &      &      &      & 7.68 & 0.45\\
    &$UBC~462^{\star}$& 7.6  & 0    &      &      &      &      & 7.68 & 0.45\\
5   &Collinder 135    & 7.42 & 0.01 &      &      & 7.60 & 0.13 & 7.42 & 0.01\\
    &UBC 7            & 7.5  & 0.1  &      &      &      &      & 7.5  & 0.1\\
6   &NGC 2659         & 7.64 & 1.21 & 6.89 & 1.59 & 7.36 & 1.58 & 7.64 & 1.21\\
    &UBC 482          & 7.43 & 0.88 &      &      &      &      & 7.43 & 0.88\\
7   &FSR 0198         & 6.67 & 2.49 & 7.0  & 2.98 & 6.56 & 2.77 & 6.67 & 2.49\\
    &Teutsch 8        & 6.6  & 2.2  & 7.0  & 1.80 & 7.50 & 1.68 & 6.6  & 2.2\\
8   &Collinder 394    & 7.96 & 0.54 &      &      & 7.86 & 0.74 & 7.96 & 0.54\\
    &NGC 6716         & 7.99 & 0.43 & 7.96 & 0.68 & 7.39 & 0.71 & 7.99 & 0.43\\
9   &Alessi 43        & 7.06 & 0.99 &      &      & 7.80 & 0.71 & 7.06 & 0.99\\
    &Collinder 197    & 7.15 & 1.42 &      &      & 7.20 & 1.70 & 7.15 & 1.42\\
10  &$FSR 1297^{\star}$&      &      &      &     & 6.70 & 5.16 & 6.76 & 0.37 \\
    &NGC 2362         & 6.76 & 0.37 & 6.70 & 0.31 & 6.64 & 0.52 & 6.76 & 0.37\\
11  &Antalova~2       & 7.08 & 1.15 &      &      & 6.0  & 11.55& 7.08 & 1.15\\
    &NGC 6383         & 6.6  & 1.3  & 6.96 & 0.92 & 6.60 & 1.29 & 6.6  & 1.3\\
12  &Negueruela 1     & 7.23 & 3.39 & 6.98 & 3.87 & 6.98 & 3.87 & 7.23 & 3.39\\
    &$Teutsch~23^{\star}$& 6.75 & 2.69 & 7.0  & 2.39 & 7.10 & 2.39 & 7.23 & 3.39\\
13  &UBC 547          & 7.18 & 1.63 &      &      &      &      & 7.18 & 1.63\\
    &UBC 549          & 7.93 & 1.92 &      &      &      &      & 7.93 & 1.92\\
14  &Stock 20         & 7.28 & 1.09 & 8.34 & 0.62 & 8.34 & 0.62 & 7.28 & 1.09\\
    & UBC 410         & 7.58 & 0.97 &      &      &      &      & 7.58 & 0.97\\
\hline
\end{tabular}
\begin{list}{}{}
{\footnotesize
  \item[] \textbf{Notes} 
  Most parameters(ages and extinctions) of the fitted isochrones of the clusters given in this work are from CG20, except the clusters marked with asterisk whose isochrone fittings are reconfirmed in this work. 
  DAML02 refers to the catalog of \citet{Dias2002}, while MWSC is short for Milky Way global survey of star clusters \citep{Kharchenko2013};
  The extinction values of DAML02 and MWSC is calculated by the extinction law $Rv = Av/E(B-V)$, with RV=3.1 \citep{1989ApJ...345..245C}. 
}
\end{list}
\label{binaries3}
\end{table*}
\section{Results}\label{section:3}
\subsection{Case analysis}

In this section, we perform the analysis for the 14 candidates of binary open clusters detected in the above section. The information related to the study is presented in Figure \ref{fig:fig1}. We compare the age and extinction of the clusters with those given in previous literature \citep{Dias2002, Kharchenko2013} to examine the fitted isochrones based on the cluster members, as shown in Table \ref{binaries3}.

\subsubsection{$h$ and $ \chi $ Persei aggregate}

Pair 1 is the well-known $h$ and $ \chi $ Persei double cluster, which is characterized by an identical mean heliocentric distance of about 2.5 kpc. The position of the two clusters can be clearly distinguished and the distance between them is 21 pc. They may be in the same stellar evolution stage, as illustrated in the first CMD panel in Figure \ref{fig:fig1}.

The aggregate is selected on the basis of transitivity, that is to say: if cluster $A$ and cluster $B$ are all friends of cluster $C$, then all the three clusters form an aggregate. We can see in Figure \ref{fig:fig1} that the members of each cluster in Pair 1 and 2 are well mixed with the members of its partner in panels a, b, c. In contrast, the members of each cluster of Pair 3 are clearly different in panel a, but well mixed in panels b, c. Nevertheless, the proper motions of these three pairs are consistent within the error range in Table \ref{binaries2}. This means that the aggregate have almost the same ages, distances, and proper motions. Using the isochrones of $h$ and $ \chi $ Persei based on the ages and extinctions given in CG20, we find that they can be well fitted with the members of clusters in Pairs 2 and 3. The derived ages for UBC46 and UBC 192 in this work are listed in Table \ref{binaries3}, showing that they are clearly younger than the ages given by CG20. Therefore, Pairs 1, 2, and 3 might form a young quadruple aggregate of an age less than 20Myr. 

\begin{figure}[!]
\centering
\includegraphics[width=\columnwidth]{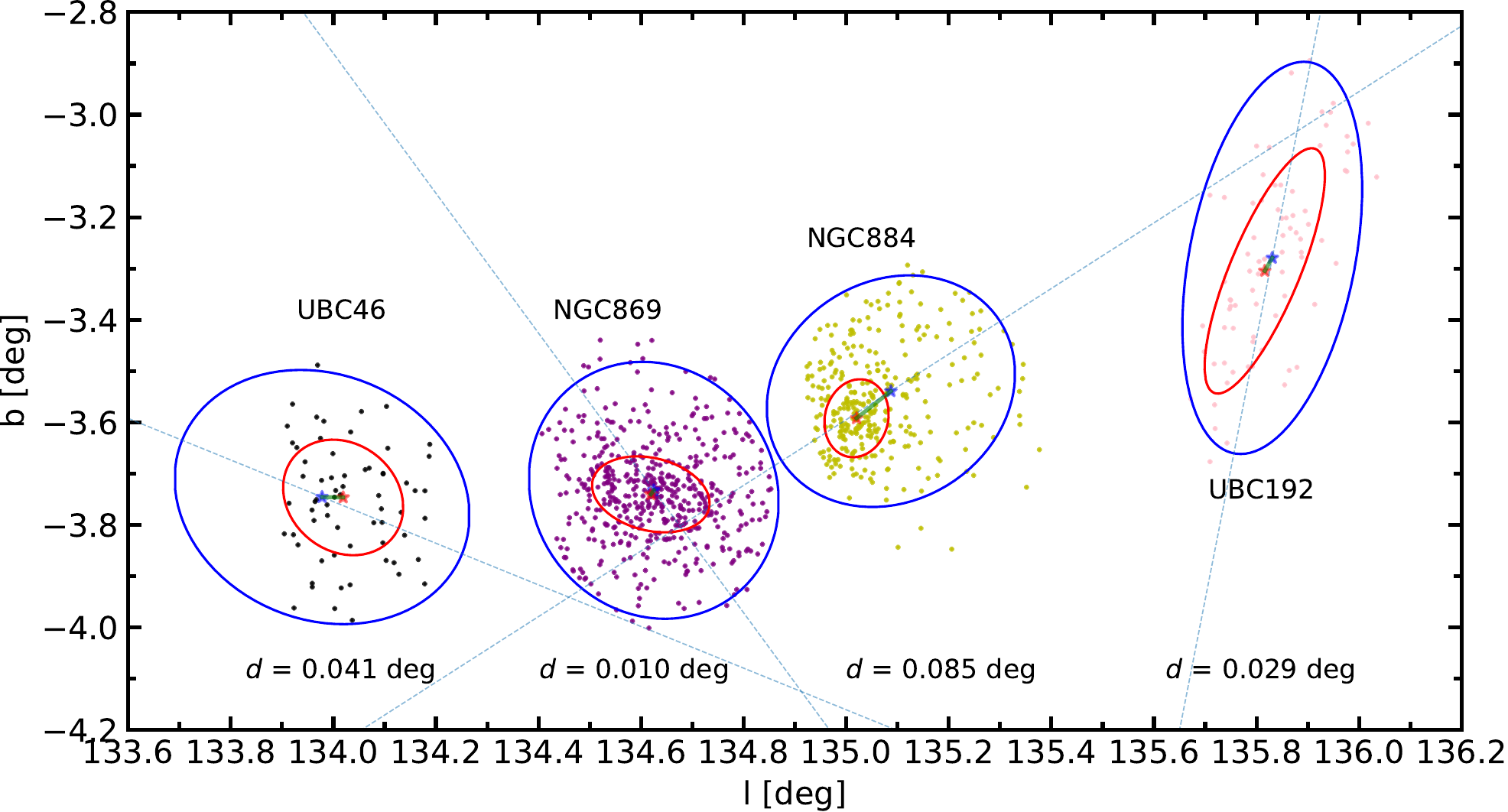}
\caption{ Distributions of the member stars (dots) of clusters in the 2D spherical Galactic coordinate system for four open clusters (UBC 46, NGC 869, NGC884, and UBC 192). The red and blue fitted ellipses are assigned as in 50\% and 5\% of the peak stellar number density of each cluster, with the red and blue asterisks being their centers, respectively. The solid green line in each cluster is the distance between the centers of inner and outer fitted ellipses regarded to the dislocation of cluster morphology \citep{2021ApJ...912....5H, 2021A&A...656A..49H}, namely, the stability of the shape of clusters. The corresponding blue dashed lines are coincident with the directions of the semi-major axes of the outer ellipses, which indicate the elongated directions of clusters.
        }
\label{fig:fig2}
\end{figure}

The distributions and movement trends of these quadruple open clusters are shown in Figure \ref{fig:fig2}. The red and blue fitted ellipses represent 50\% and 5\% of the peak stellar number density profiles for each cluster, with the red and blue asterisks being their centers, respectively. The blue dashed lines are coincident with the semi-major axes of the outer ellipses, indicating the direction of the cluster elongations \citep{2021ApJ...912....5H}. The red and blue pentagrams of each cluster are linked with a solid green line, which is the distance between the centers of inner and outer fitted ellipses, also regarded as the dislocation of cluster morphology initially defined by \cite{2021A&A...656A..49H}. In their work, the morphological dislocation of an open cluster refers specifically to the Euclidean distance between the centroids of the inner and outer fitted ellipses of the open cluster in the 2D spherical coordinate system. The morphological dislocation measures the stability of the 2D morphology of the cluster; and the larger the morphological dislocation, the less stable the 2D morphology of the cluster. 

We find that the three clusters of this group (UBC 46, NGC 869, and NGC 884) seem to be elongated as a result of their interaction with each other, since the shaped direction of their outer ellipses intersects with each other. NGC 869 may be the central cluster which are likely to disturb or influence the shapes of UBC 46 and NGC 884; this is because it has a greater number of member stars than the other two clusters. Generally speaking, the higher the number of member stars, the stronger the self-gravity of a cluster and the stronger the cluster’s gravity is expected to be. Moreover, in comparing their dislocations of morphologies, we can note the smallest dislocation of shape in NGC 869, implying the morphological stability of the cluster is higher than that of the rest clusters. This reinforces the idea of NGC 869 as the central cluster of the group. UBC 46 and NGC 884 have dislocations of morphologies about four and eight times that of NGC 869, respectively, which means that their morphological stabilities may be smaller than NGC 869 and more influenced by the central cluster. In addition, although UBC 192 displays an elongated shape that is even larger than the three other clusters and its small dislocation of morphology seems to suggest a high probability with regard to the morphological stability.

\citet{2021ApJ...909...90D} found a stellar complex known as LISCA I, which is composed of seven comoving clusters with an extended massive halo distributed within 6 degrees surrounding the well-known $h$ and $ \chi $ Persei double stellar cluster in the Perseus Arm by Gaia-EDR3 data. The stellar complex share the same chemical abundances of half solar metallicity and the same age of $t \sim 20 Myr$, which could  place them at a distance of 2.3 kpc and the extinction value set at Av = 1.65 mag. The values of the parameters are consistent with our results. Our findings extend the stellar complex and the new members of UBC 46 and UBC 192 are much closer than the others (the cluster separations are 29.30 pc and 38.59 pc, respectively).
 
\subsubsection{Known binary open clusters}

The cluster pair made up of UBC~461 and UBC~462, both open clusters discovered by \citet{2020A&A...635A..45C}, was first mentioned in \citet{2021ARep...65..755C} as a probable physical system of three OC aggregates. Their ages and extinctions  are provided by CG20. The PARSEC track fits in well with the cluster parameters of UBC~461. However, for UBC~462, the isochrones based on the adopted cluster parameter cannot fit the color-magnitude diagram constituted by cluster members, due to the fact that only eight member stars are available for this cluster. Their indistinguishable CMDs allow them to share the same track of UBC~461 as shown in Figure \ref{fig:fig1}, which means that the age of UBC~462 may be comparable to that of UBC~461. 
    
The cluster pair Collinder~135 and UBC~7 was identified as a physical pair that has since been analyzed in detail by \citet{2020A&A...642L...4K,2021scgr.confE..17K}. In addition, \citet{2020A&A...642L...4K} built a model to show that the two clusters might indeed form a physical pair. Collinder~135 is a well-known open cluster and UBC~7 used to be considered as part of Collinder~135 before Gaia Data was released \citep{2020A&A...642L...4K, Kharchenko2013}. This pair is the only pair in our samples where the clusters are distinctly separated by their parallax histograms, rather than by equatorial coordinates. This indicates that they are obviously separated along  the line of sight.

NGC~2659 is a young open cluster that is characterized the cluster parameters given by \citet{Kharchenko2013}. UBC~482 was discovered by the work of \citet{2020A&A...635A..45C}. The cluster pair NGC~2659 and UBC~482 are another pair discovered by \citet{2021ARep...65..755C}, which makes up part of a large group containing nine members: Casado 28, NGC 2659, Casado 61, NGC 2645, SAI 92, LP 58, Pismis 8, UBC 482, and Gulliver 5. With the cluster spatial separation of 49.45 pc, this pair has the largest gap among our selected pairs. 

The cluster pair FSR~0198 and Teutsch~8 can be found in the works of \citet{2021A&A...649A..54P} and \citet{2022Univ....8..113C}, respectively. They are likely members of a triple primordial group of FSR~0198, Teutsch~8, and NGC~6871 \citep{2022Univ....8..113C}; later, Biurakan~2 was added to the group in \citet{2021A&A...649A..54P}. All the relevant parameters of this pair are well-matched except the celestial positions, as illustrated in Figure~\ref{fig:fig1}.

Collinder~394 and NGC~6716 were confirmed to be physically interacting binaries by \citet{2022MNRAS.510.5695A}. The pair was discovered by \citet{2019ApJS..245...32L}, and \citet{2020JPhCS1523a2013N} were the first to analyze their morphology, age estimates, photometric mass, and kinematics using Gaia DR2 for the first time. This is the oldest (age > 90 Myr) and closest (Sep = 11.49 pc) pair compared to the other clusters in this work. 

The cluster pair Alessi~43 (also named ASCC~50) and Collinder~197 was first studied by \citet{2009A&A...500L..13D}. In \citet{2019ApJS..245...32L}, it was assigned to a triple group that included Collinder~197, Alessi~43, and BH~56. However, \citet{2021A&A...649A..54P} argued that it could be a binary cluster. All the relevant parameters (apart from the celestial coordinates) are well matched, as seen in Figure~\ref{fig:fig1}.

\subsubsection{New binary open clusters}

Pairs 10-14 are newly discovered candidates of binary open clusters in this work. Each pair fits well with the parameters of CG20 (see Figure \ref{fig:fig1} and Table \ref{binaries2}). In particular, UBC~410, UBC~547, and UBC~549 were discovered by \citet{2020A&A...635A..45C}. The other clusters were previously investigated by \citet{Kharchenko2013}. In this section, we present an analysis of the new candidate pairs and compare the cluster parameters with those investigated by the previous literature \citep{Dias2002, Kharchenko2013}.

The distance between FSR~1297 and NGC~2362 is 41.13 pc. It is easy to distinguish them from the RA/DEC coordinates. Their CMDs overlap and the clusters share the same isochrone of NGC~2362, which indicates that FSR 1297 probably has the same age and extinction as NGC 2362 as shown in Figure \ref{fig:fig1}. The derived age of FSR~1297 is coincident with \citet{Kharchenko2013}.

The ages provided by \citet{Kharchenko2013} for Antalova~2 and NGC~6383 are $logt = 6.0$ yr, and $logt = 6.6$ yr, respectively. The given age for Antalova~2 is quite different from the cluster age of $logt = 7.08$ yr from CG20, as shown in Table \ref{binaries3}. From the CMD in Figure \ref{fig:fig1}, the two isochrones from CG20 both fit well with the cluster members. There are 25 member stars in Antalova~2, which are scattered among the members of NGC~6383. The low number of known member stars makes it difficult to choose a better isochrone for Antalova~2. More member stars are needed to identify the actual age of Antalova~2. However, this does not affect our conclusion that it is a candidate for a truly binary open cluster. 

The cluster separation of  Negueruela~1 and Teutsch~23 is 18.31 pc. The isochrone of Teutsch~23 based on Table \ref{binaries2} does not fit well with the cluster members. The ages given in Table \ref{binaries2} show that Negueruela~1 ($logt = 7.23$ yr) is about 11 Myr older than Teutsch~23 ($logt = 6.75$ yr), which is contradiction to the works of \citet{Kharchenko2013} and \citet{Dias2002}, where the ages in $logt$ for Negueruela 1 and Teutsch 23 are 6.98 yr and 7.0 yr, respectively (as shown in Table \ref{binaries3}). The isochrone of Negueruela~1 fits in well with the member stars of Teutsch~23, indicating that the ages of the two clusters are almost the same, which is consistent with the literature.

UBC~547 and UBC~549 were discovered to be open clusters by \citet{2020A&A...635A..45C}. The member stars are mixed in the proper motion space, color-magnitude diagram, and parallax histogram. The derived isochrones by CG20 could potentially fit in with the members, but they are not adequate for a physical binary cluster. The age of UBC~549 (85 Myr) given by CG20 has to be considered with caution since there are no evolved member stars at the upper part of CMD. If this is a true pair, the age of UBC~549 would be comparable to that of UBC~547.

The cluster parameters of candidate cluster pair Stock~20 and UBC~410  fit in well to their cluster members, as shown in Figure \ref{fig:fig1}. The age of Stock~20 is specified as $logt = 8.337$ yr, provided by \citet{Kharchenko2013} and \citet{Dias2002}. In CG20, the age of Stock~20 decreases to $logt = 7.28$ yr, which is more suitable for the cluster members. The cluster separation is 23.74 pc.

Based on our analysis of the candidates of truly pairs open clusters, the cluster distances in the proper motion space are limited to 0.86 mas/yr, and the parallax differences between the clusters are less than 0.28 mas. If ignoring the disputed pairs, the ages of all candidates of pair OCs are < 100 Myr, with the age gaps less than 20 Myr, and the extinction discrepancies are limited to 0.4 mag. The ages of most binary open clusters can further decrease to 50 Myr, except for the oldest pair, Collinder~394 (91.2 Myr) and NGC~6716 (97.72 Myr). Our results are consistent with the literature \citep{1990PASJ...42..757B, 2022Univ....8..113C}, confirming that the binary open clusters are composed of young star clusters. 

\subsection{Comparison with 2MASS}

\begin{figure*}[!]
   \centering
   \begin{tabular}{ccc}
   \subfloat{\includegraphics[scale=0.32]{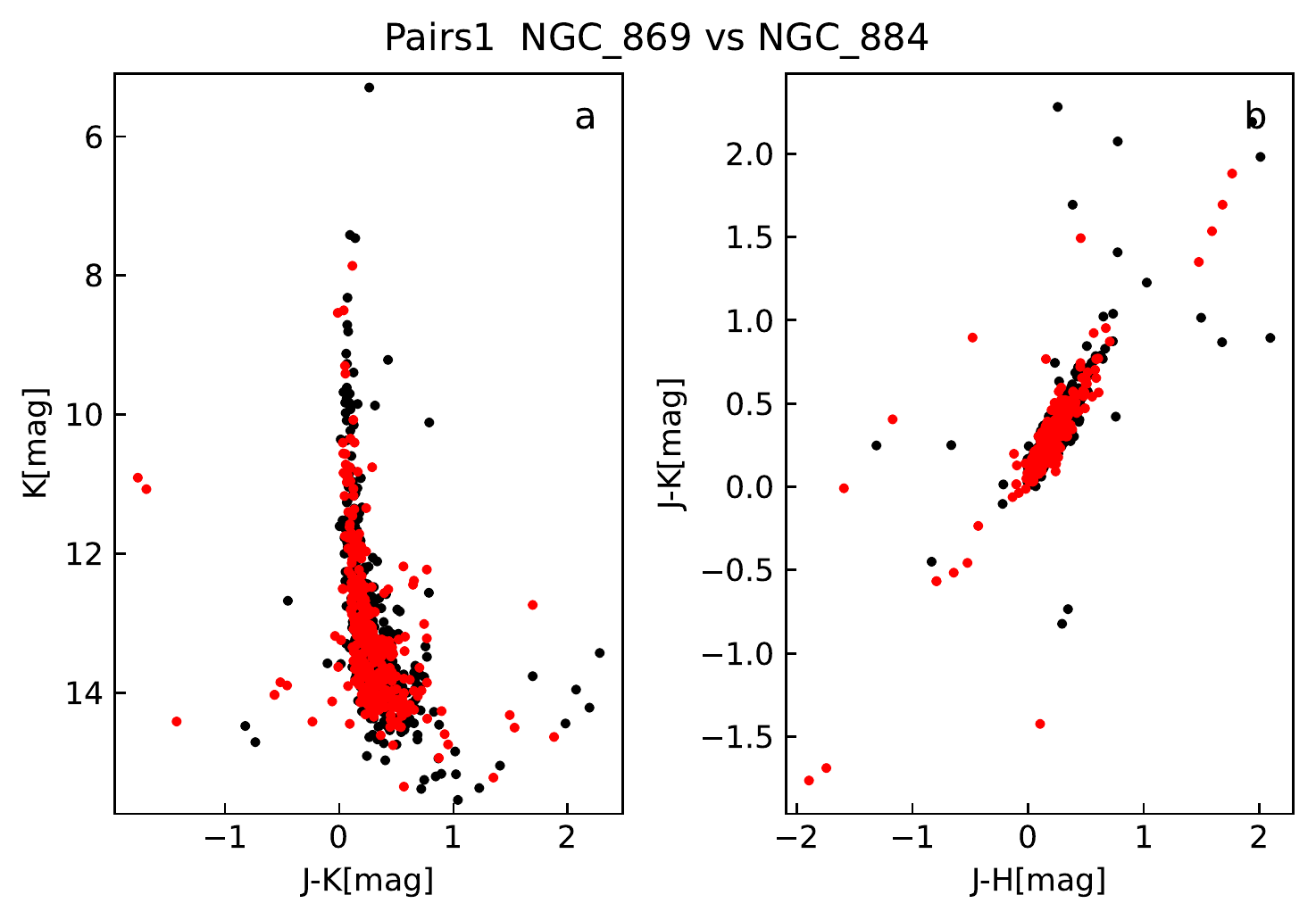}} \quad &   
   \subfloat{\includegraphics[scale=0.32]{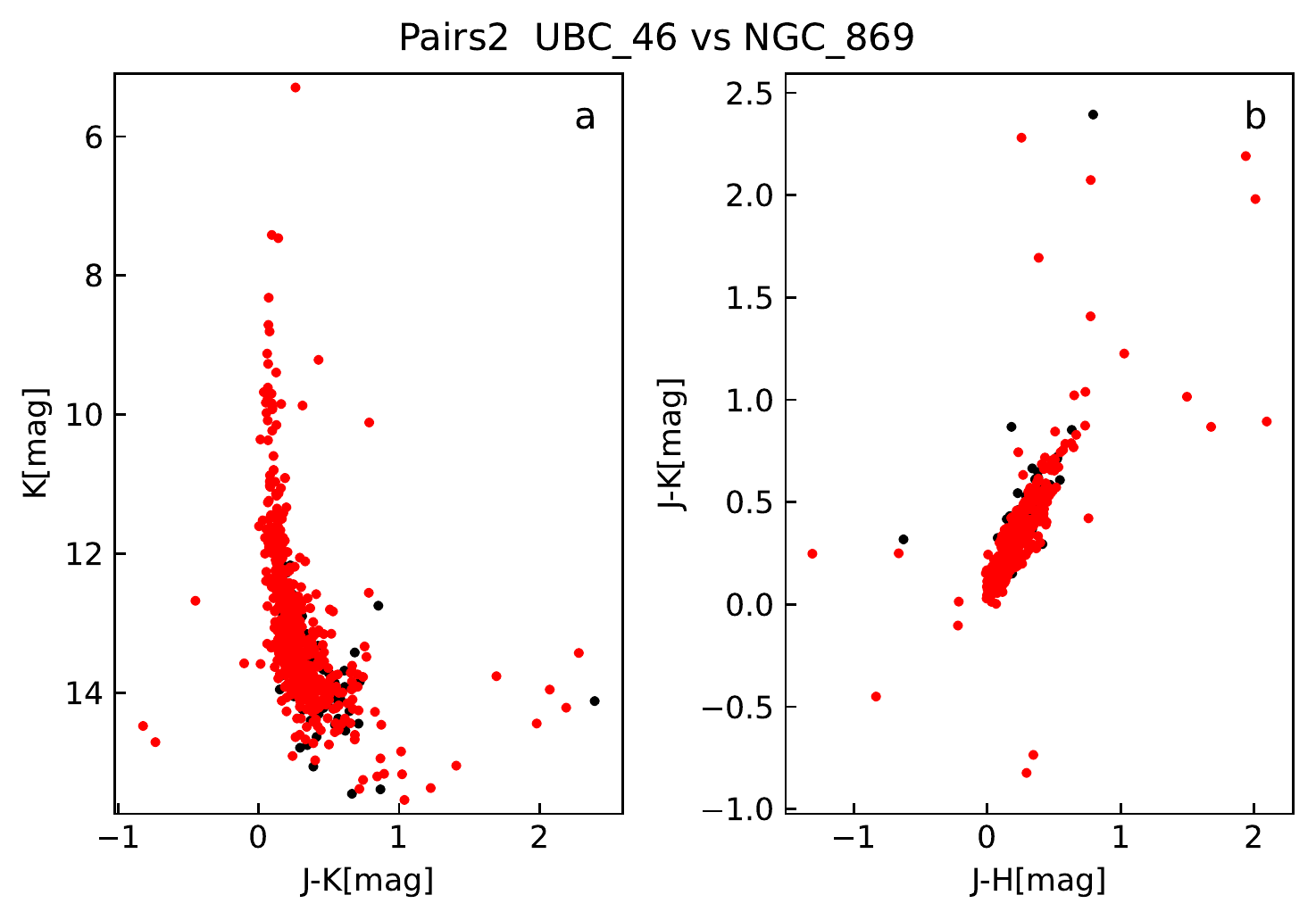}} \quad &
   \subfloat{\includegraphics[scale=0.32]{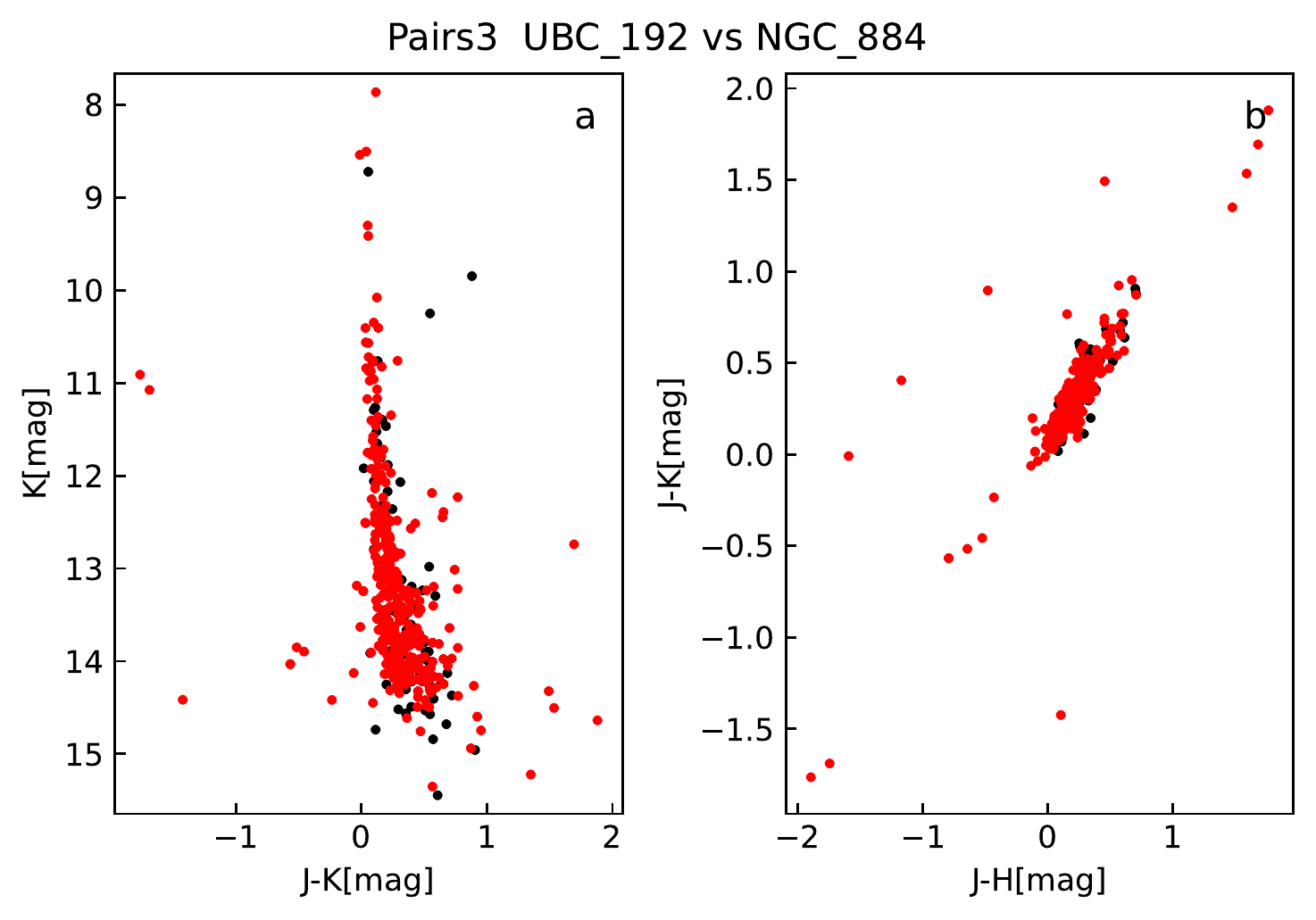}} \\ 
   \subfloat{\includegraphics[scale=0.32]{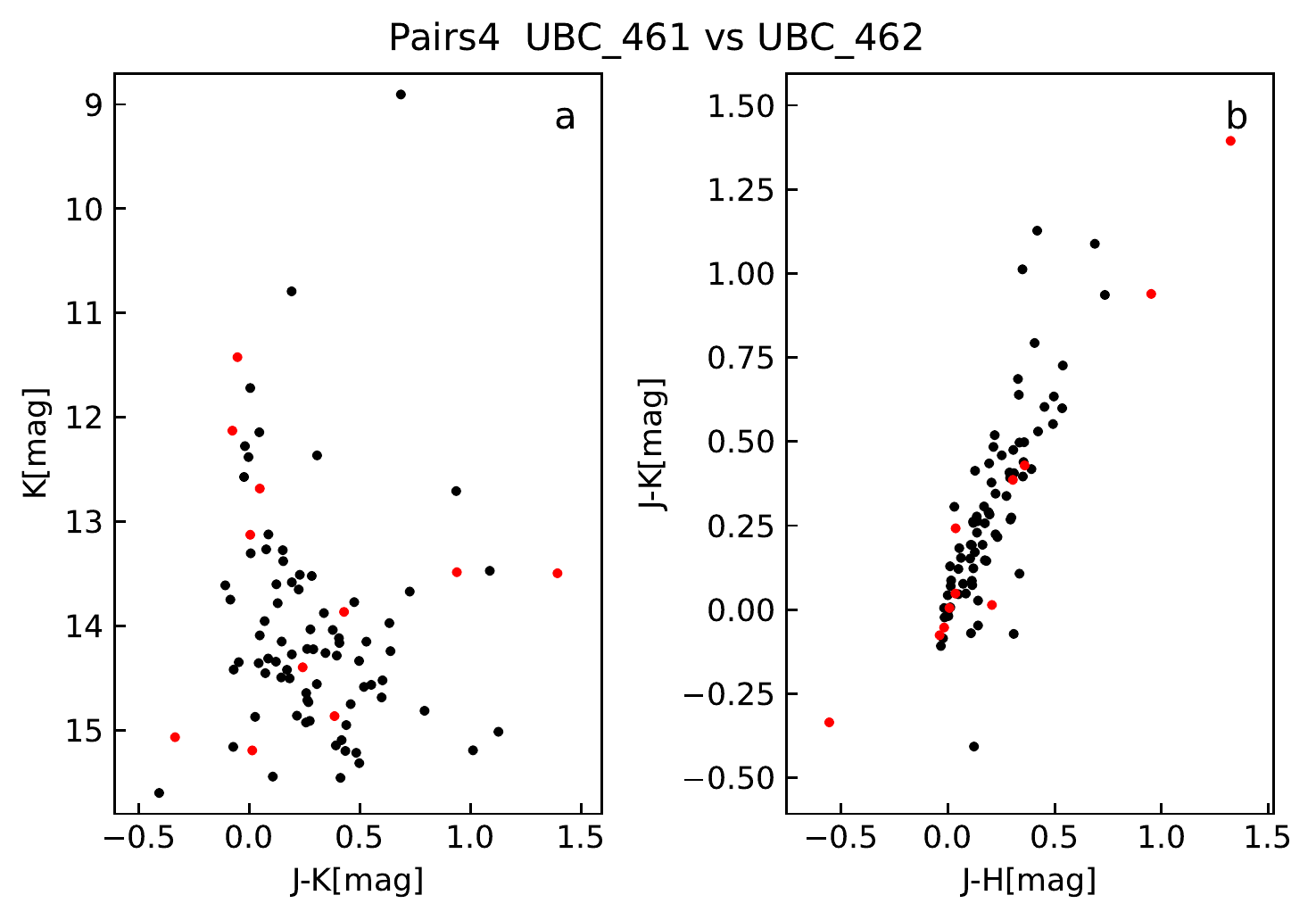}} \quad &
   \subfloat{\includegraphics[scale=0.32]{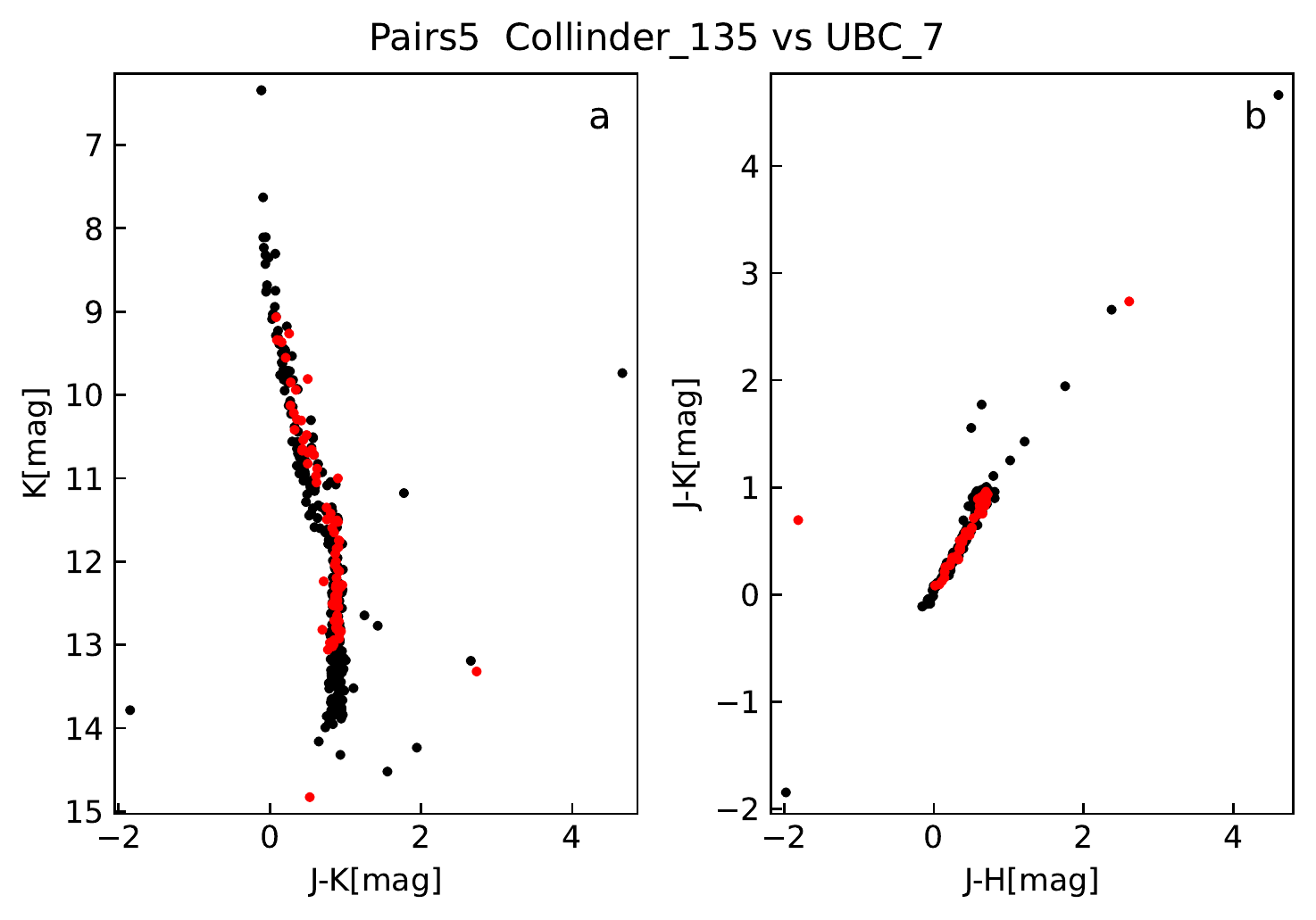}} \quad & 
   \subfloat{\includegraphics[scale=0.32]{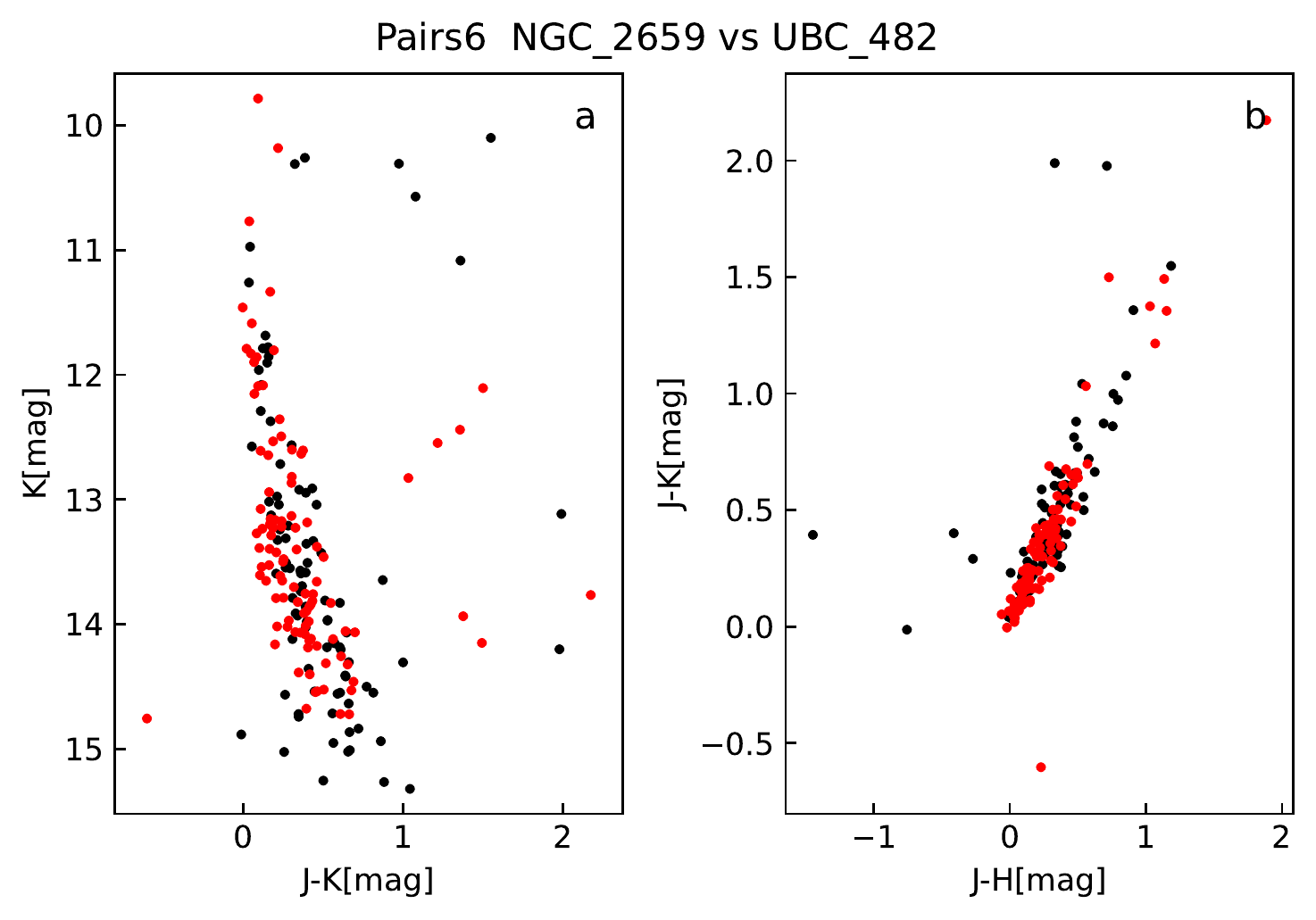}} \\
   \subfloat{\includegraphics[scale=0.32]{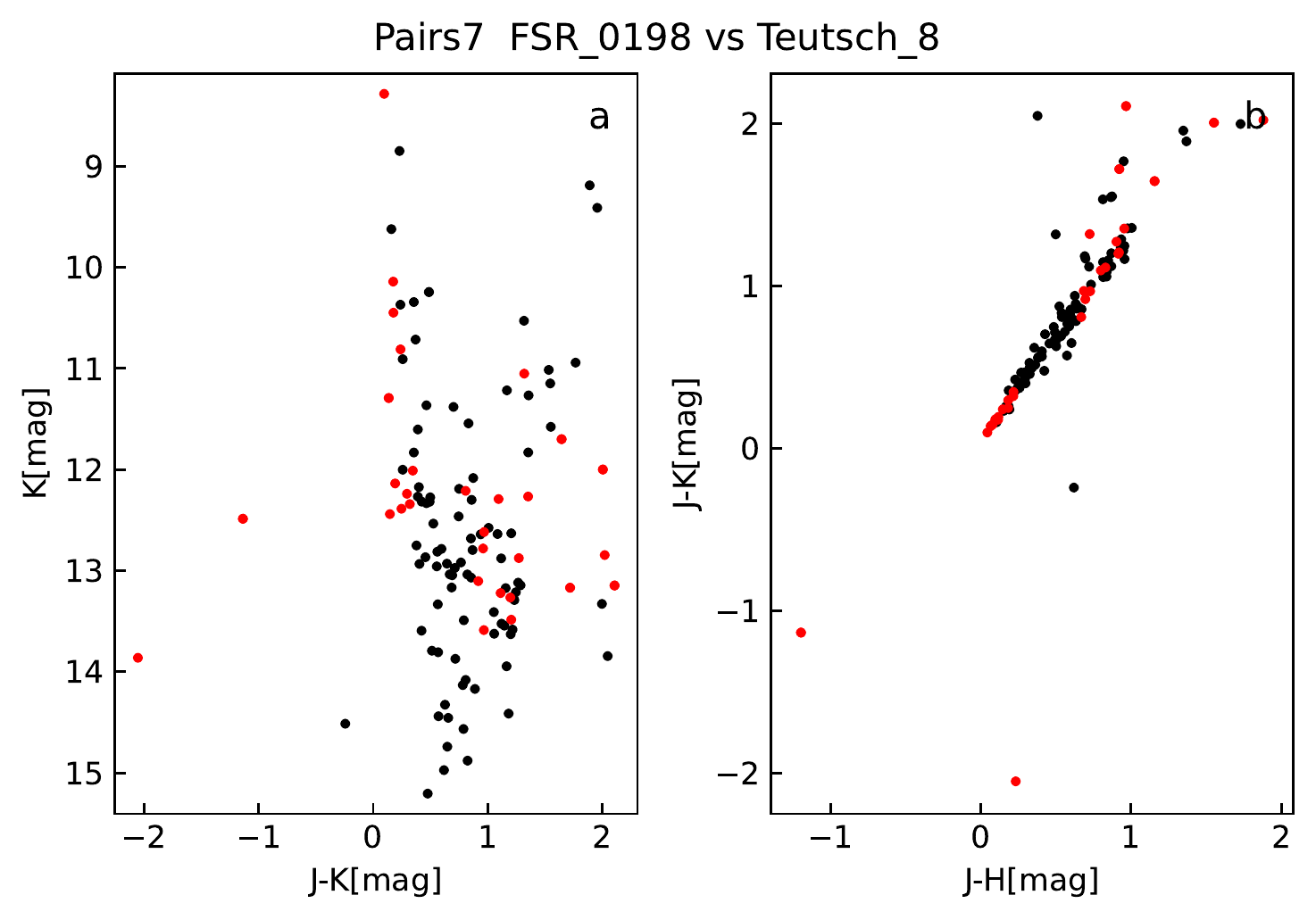}} \quad & 
   \subfloat{\includegraphics[scale=0.32]{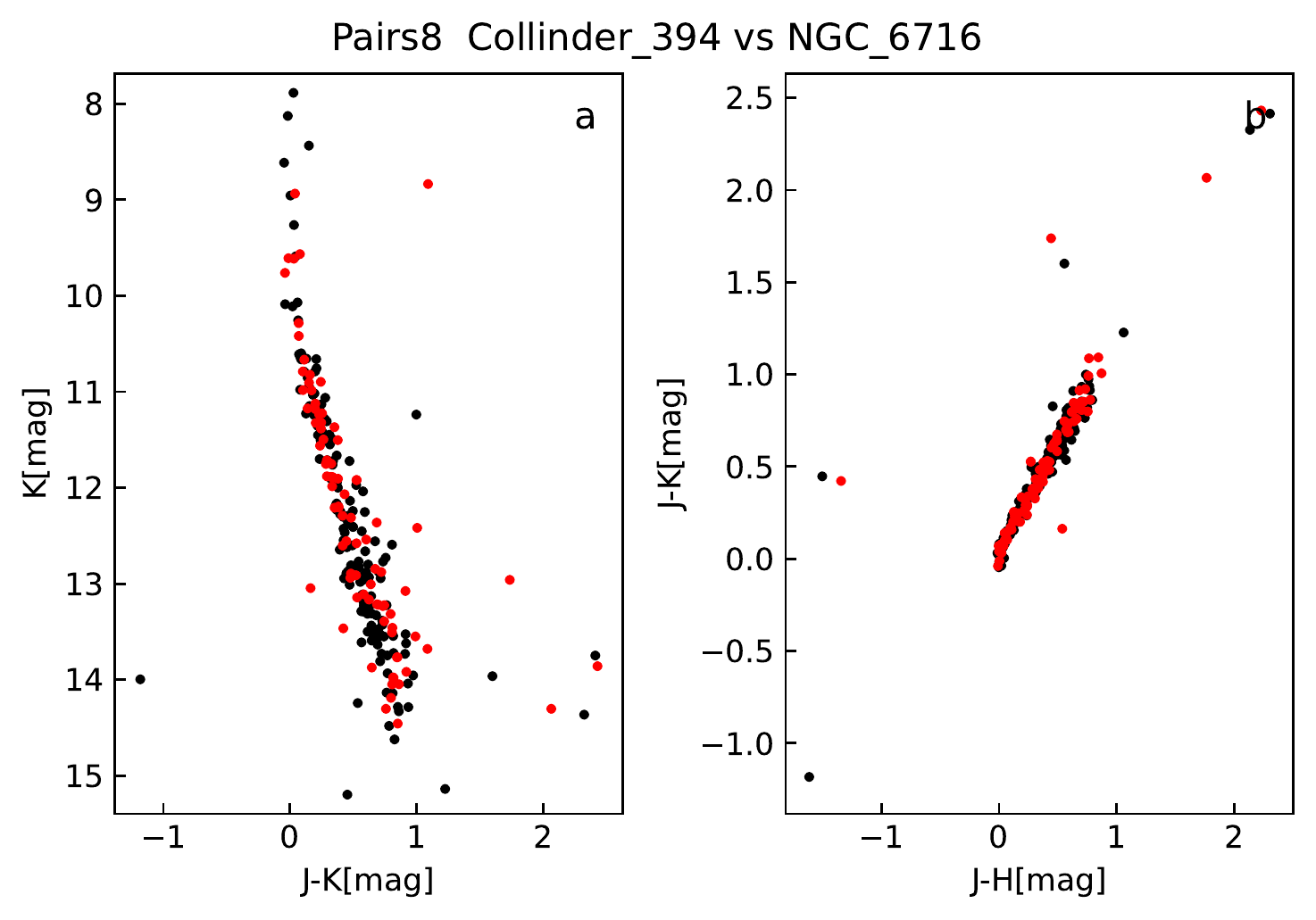}} \quad &
   \subfloat{\includegraphics[scale=0.32]{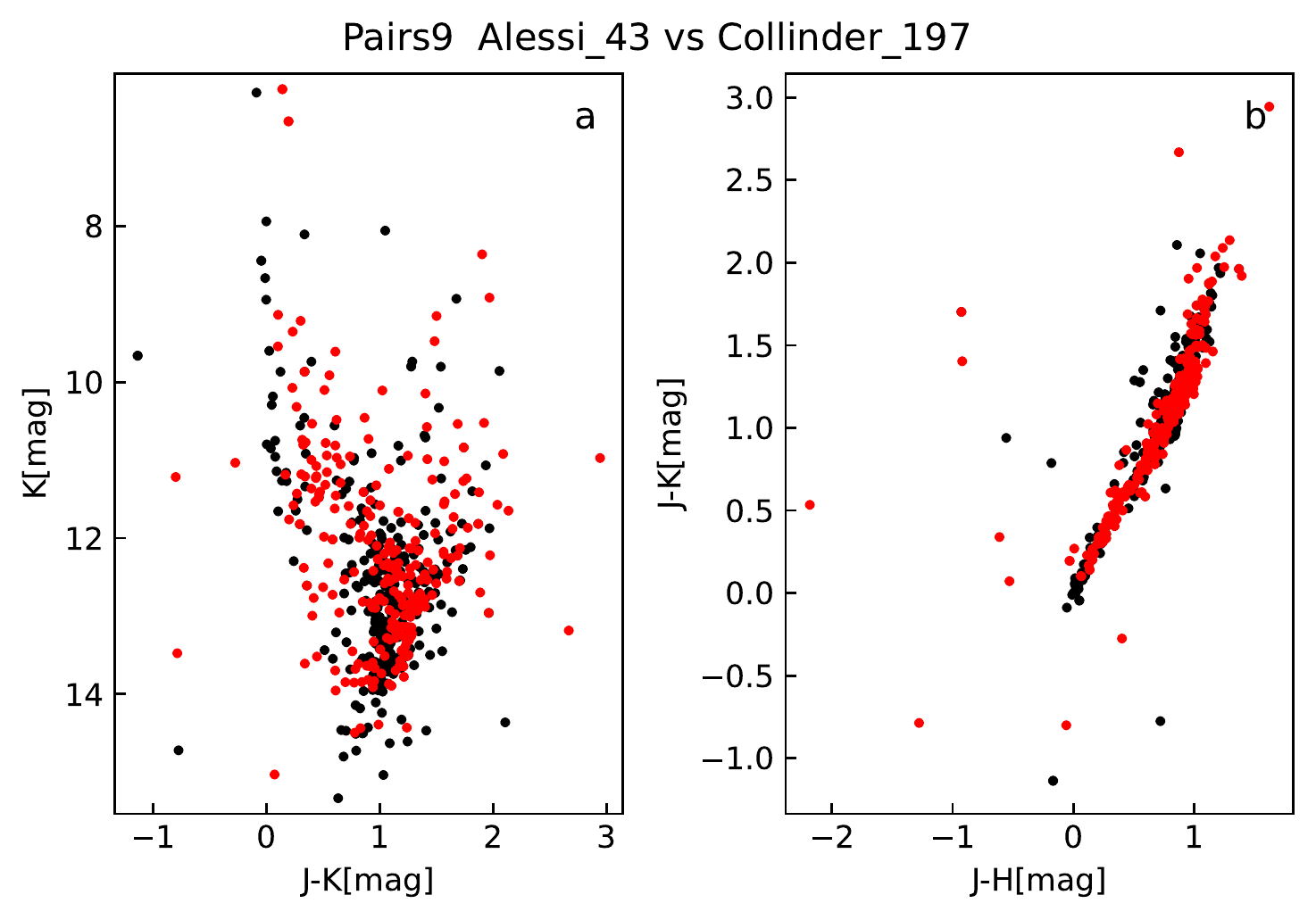}}  \\
   \subfloat{\includegraphics[scale=0.32]{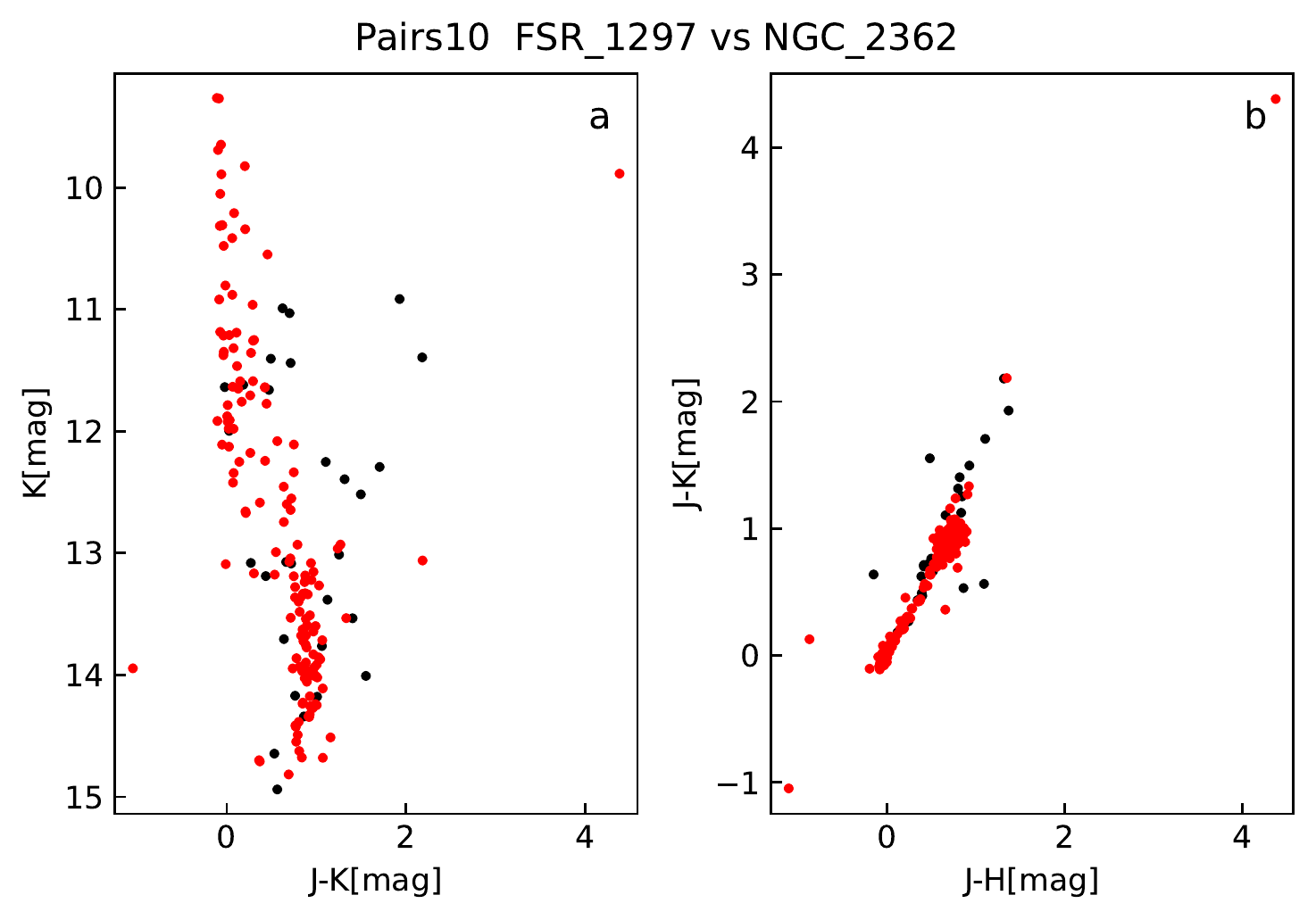}} \quad &
   \subfloat{\includegraphics[scale=0.32]{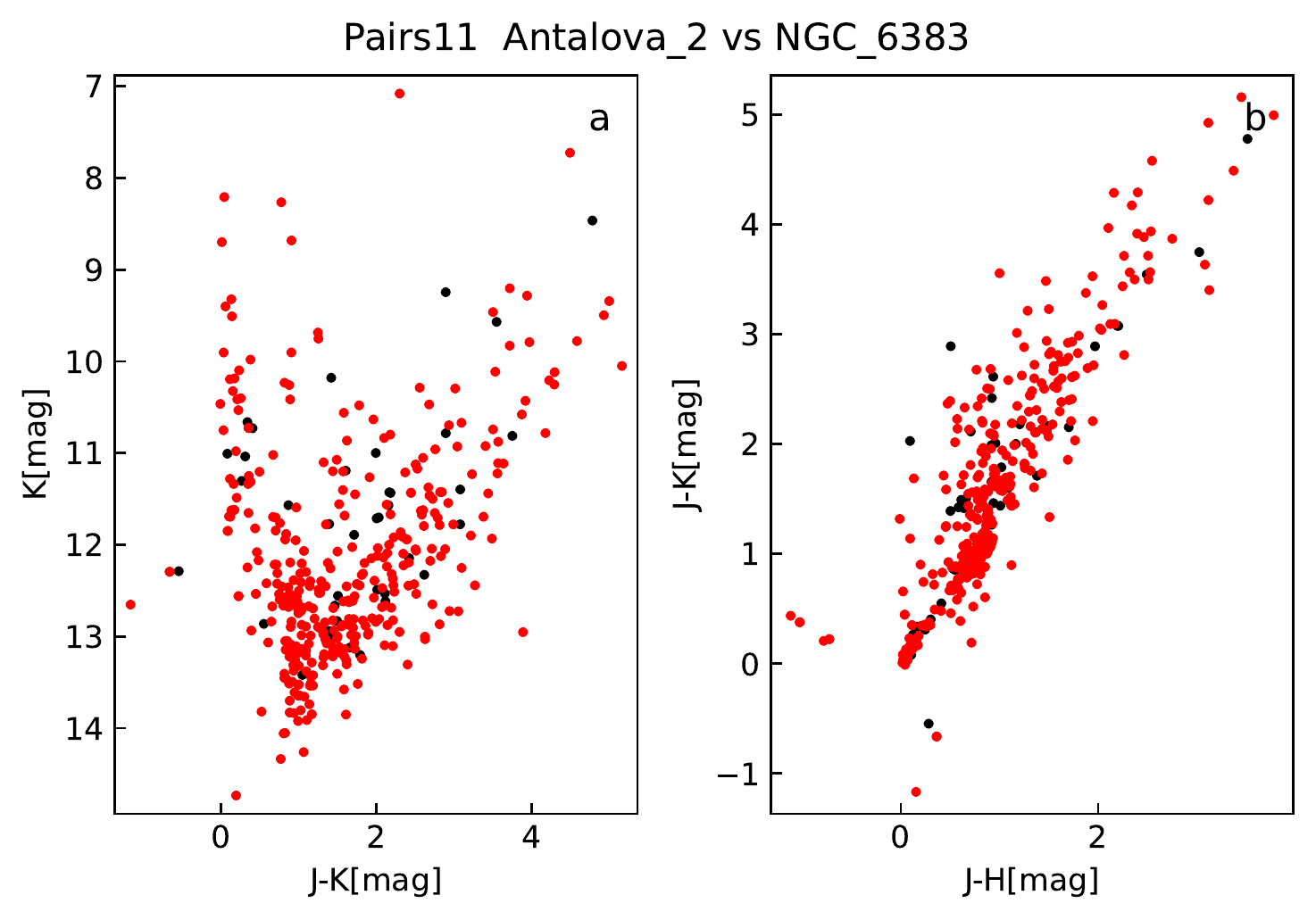}} \quad & 
   \subfloat{\includegraphics[scale=0.32]{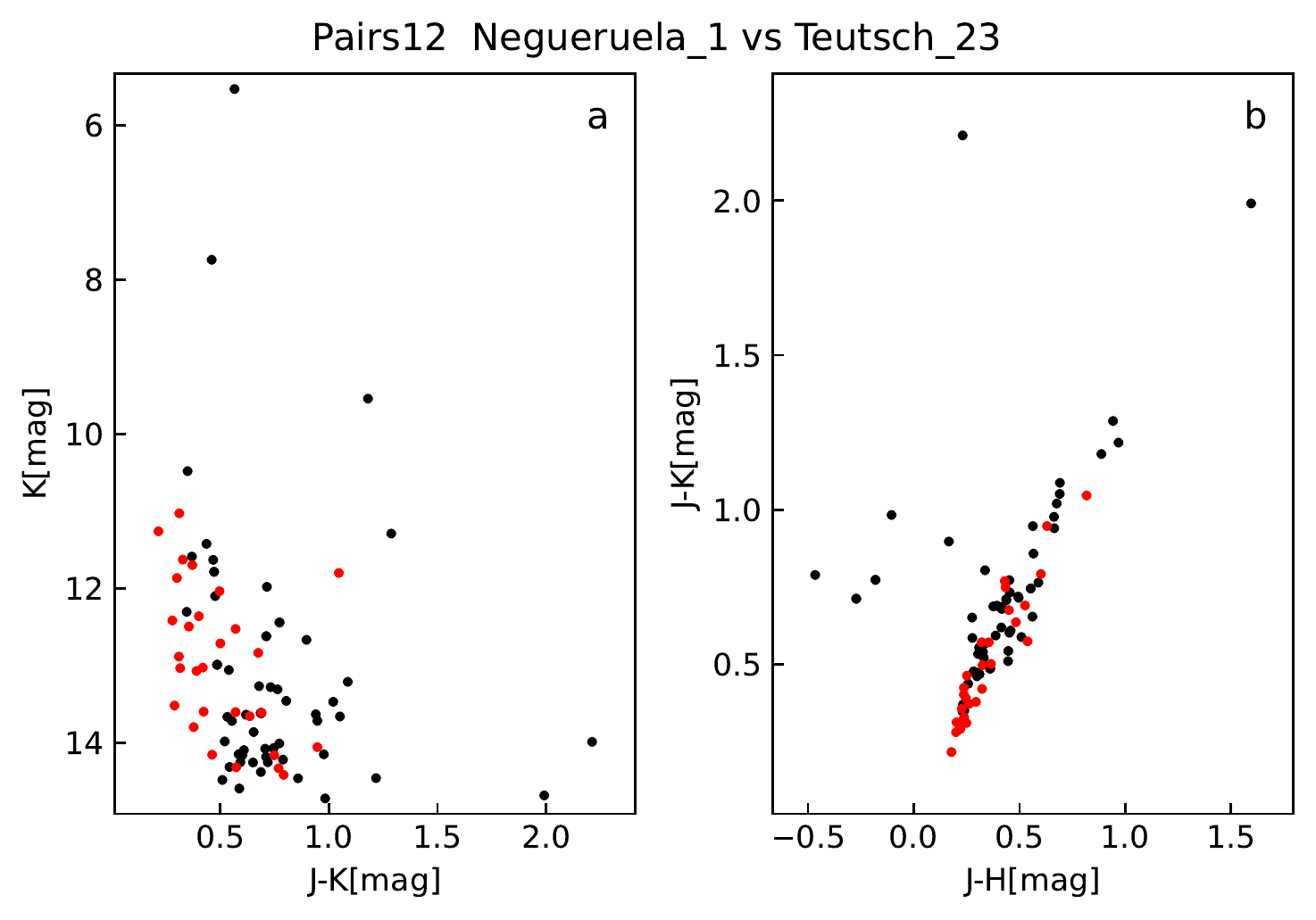}} \\
   \subfloat{\includegraphics[scale=0.32]{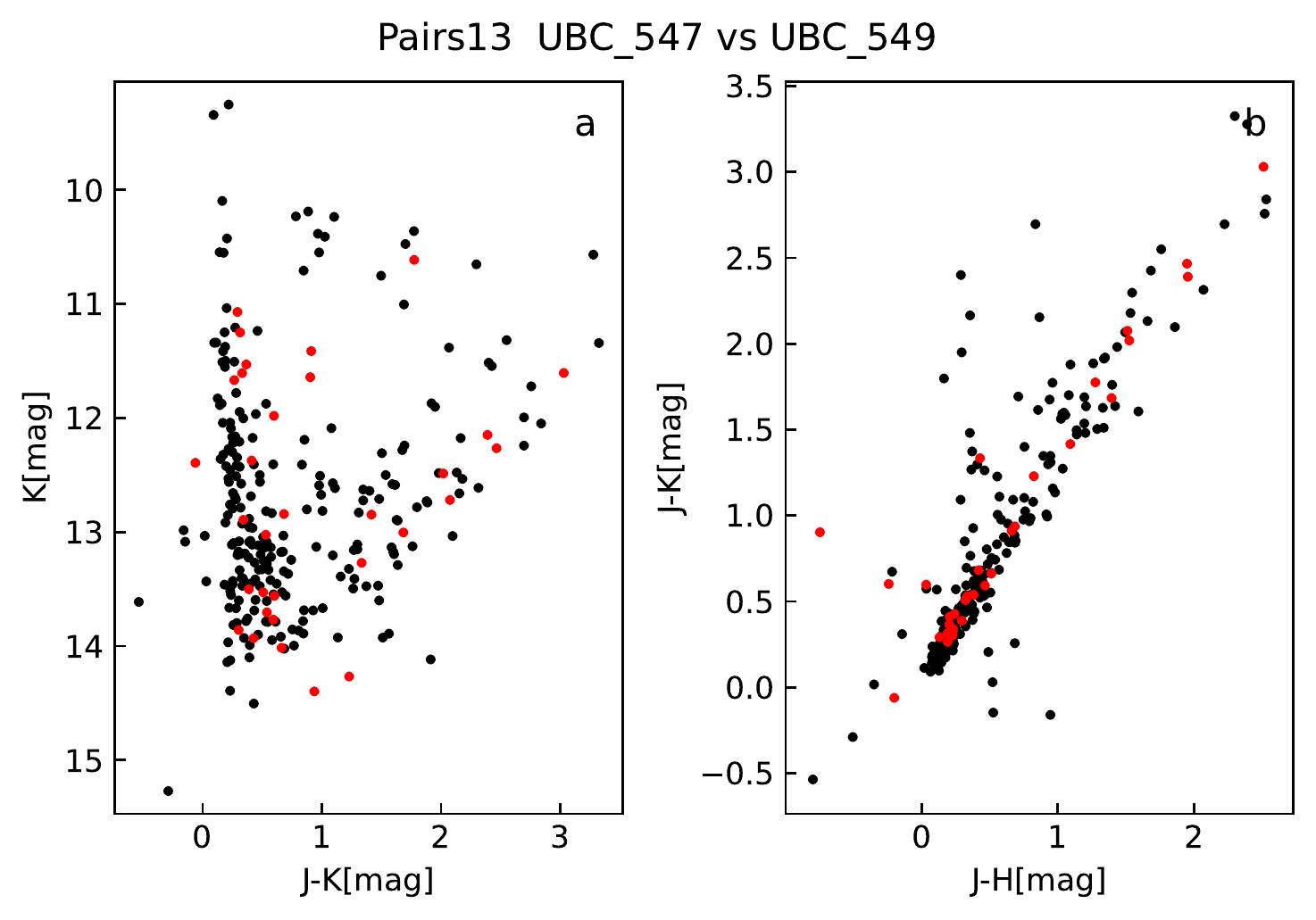}} \quad & 
   \subfloat{\includegraphics[scale=0.32]{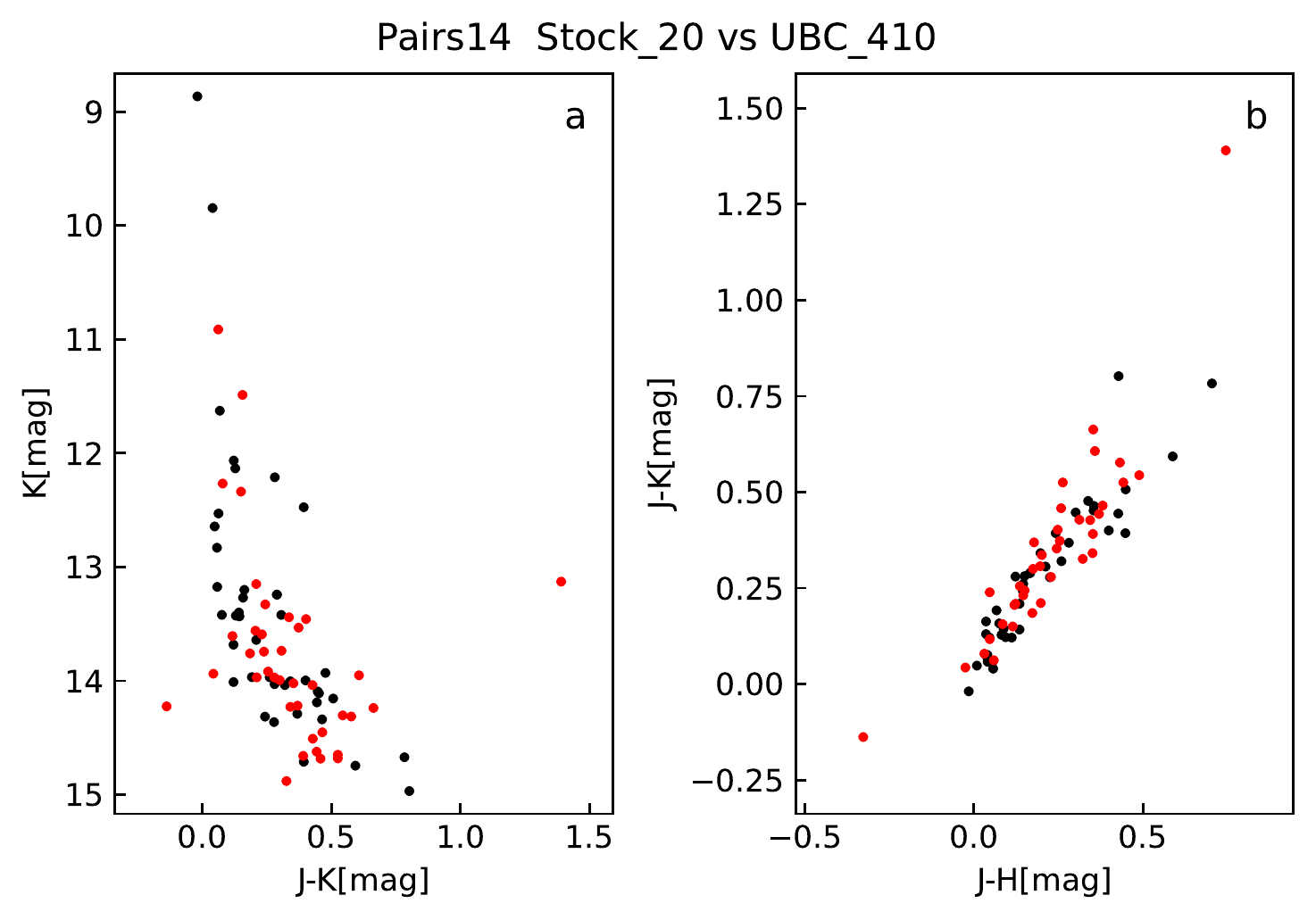}} \quad &  \\
         
   \end{tabular}
   \caption{ Distributions of colour-magnitude diagrams and color-color diagrams of members of 14 open cluster pairs using the 2MASS photometric data. For each pair, plot (a) represents the colour-magnitude diagram and plot (b) gives the color-color diagrams. The dark dots represent members of the former clusters and the red dots signify members of the latter clusters.}
   \label{fig:fig3}
\end{figure*}

The Two Micron All Sky Survey (2MASS) catalog \citep{2003yCat.2246....0C} provides high-quality near-infrared photometry for all-sky stars. We searched for the stars of the 14 analyzed pairs in 2MASS catalog based on the coordinates of the member stars. Figure \ref{fig:fig3} presents the distributions of color-magnitude diagrams (CMDs) and color-color diagrams (CCDs) for members of the 14 open cluster pairs using 2MASS photometric data. As shown in Figure \ref{fig:fig3}, all the pairs do not show a distinction between the two constituent clusters.

\section{Summary}\label{section:4}

In this study, we identify14 candidates of truly binary open clusters (see Table \ref{binaries2}), including one aggregate (Pairs 1-3, composed of four clusters: the $h$ and $ \chi $ Persei double cluster, UBC 46, and UBC 192), based on the Gaia database and cluster separations limited to 50 pc. By analyzing the distributions and movement trends of these quadruple open clusters, we find that NGC 869 may be the central cluster, with enough potential to disturb or influence the shapes of UBC 46 and NGC 884. Moreover, the morphological stability of NGC 869 is higher than that of the remaining clusters. Pairs 4-9 are known binary open clusters that can be found in the literature, and the parameters we obtained are essentially consistent with the literature.  Pairs 10-14 are newly identified candidates of binary open cluster pairs, which are first reported in this work. Incidentally, ten pairs (Table \ref{binaries1}) appear to be the same clusters based on the fact that almost half of the cluster members are shared with one another. Among them, four pairs with ratio values greater than 0.8 have a greater probability of being the same clusters, while the other pairs require further study. 

\begin{acknowledgements}
We are grateful to an anonymous referee for valuable comments which have improved the paper significantly. This work has been financially supported by the Natural Science Foundation of China (NSFC-U2031209). This work has made use of data from the European Space Agency (ESA) mission GAIA processed by Gaia Data processing  and Analysis Consortium (DPAC), (https://www.cosmos.esa.int/web/gaia/dpac/consortium).
\end{acknowledgements}

\bibliographystyle{aa} 
\bibliography{aanda}

\end{document}